\documentclass[reprint,twocolumn,aps,prb,10pt,amsmath,amssymb,longbibliography,superscriptaddress,
    showpacs,preprintnumbers,nofootinbib]{revtex4-2}

\usepackage{graphicx}%
\usepackage{siunitx}
\usepackage{bm}
\usepackage[caption=false]{subfig}
\usepackage{mathtools}
\usepackage{multirow}
\usepackage{amssymb}
\usepackage{braket}
\usepackage{amsmath}
\usepackage{placeins}
\usepackage{cancel}
\usepackage{xcolor}
\usepackage{physics}
\usepackage{spreadtab}
\usepackage{dsfont}
\usepackage[normalem]{ulem}  
\usepackage[colorlinks=true,
            linkcolor=blue,
	        citecolor=blue,
	         urlcolor=blue]{hyperref}

\begin{document}

\title{Electron-phonon coupling in magnetic materials using the local spin density approximation}

\author{\'Alvaro Adri\'an Carrasco \'Alvarez}
\email{alvaro.carrasco@uclouvain.be}
\affiliation{European Theoretical Spectroscopy Facility, Institute of Condensed Matter and Nanosciences, Université catholique de Louvain, Chemin des Étoiles 8, B-1348 Louvain-la-Neuve, Belgium}
\author{Matteo Giantomassi}%
\affiliation{European Theoretical Spectroscopy Facility, Institute of Condensed Matter and Nanosciences, Université catholique de Louvain, Chemin des Étoiles 8, B-1348 Louvain-la-Neuve, Belgium}
\author{Jae-Mo Lihm}%
\affiliation{European Theoretical Spectroscopy Facility, Institute of Condensed Matter and Nanosciences, Université catholique de Louvain, Chemin des Étoiles 8, B-1348 Louvain-la-Neuve, Belgium}
\author{Guillaume E. Allemand}
\affiliation{European Theoretical Spectroscopy Facility, Nanomat/Q-Mat Universit\'e de Liège (B5), B-4000 , Liège Belgium}
\author{Maxime Mignolet}
\affiliation{European Theoretical Spectroscopy Facility, Nanomat Q-Mat University of Liège}
\author{Matthieu Verstraete}%
\affiliation{European Theoretical Spectroscopy Facility, Nanomat Q-Mat University of Liège}
\affiliation{ITP Dept of Physics, University of Utrecht, 3508 TA Utrecht, The Netherlands}
\author{Samuel Ponc\'e}%
\email{samuel.ponce@uclouvain.be}
\affiliation{European Theoretical Spectroscopy Facility, Institute of Condensed Matter and Nanosciences, Université catholique de Louvain, Chemin des Étoiles 8, B-1348 Louvain-la-Neuve, Belgium}
\affiliation{WEL Research Institute, avenue Pasteur 6, 1300 Wavre, Belgium.}

\date{\today}

\begin{abstract}
Magnetic materials are crucial for manipulating electron spin and magnetic fields, enabling applications in data storage, spintronics, charge transport, and energy conversion, while also providing insight into fundamental quantum phenomena.
In numerous applications, the interaction between electrons and lattice vibrations, known as electron-phonon coupling, can be of significant importance. 
In that regard, we extend the \textsc{EPW} package to be able to interpolate the electron-phonon matrix elements combining perturbation theory and maximally localized Wannier functions.
This enables the use of dense momentum grids at a reasonable computational cost when computing electron-phonon-related quantities and physical properties.
We validate our implementation considering ferromagnetic iron and nickel, where we explore the phonon induced mass enhancement and Eliashberg spectral function finding different importance of each spin channel for both compounds. 
Furthermore, we evaluate the carrier resistivity at finite temperatures for both systems, considering the role of the magnetic phase in carrier transport. 
Our findings indicate that in the case of Fe, the primary contributor to resistivity is electron-phonon scattering. 
In contrast, for Ni, electron-phonon scattering constitutes less than one-third of the resistivity, underscoring a fundamental difference in the transport properties of the two systems.
\end{abstract}

\maketitle

\section{Introduction}\label{sec:level1}

Magnetic materials play an important role for technological applications, including magnetic cores in electric motors and generators, or magnetic memory devices in computers and mobile phones. 
From a historical standpoint, iron is arguably the most significant magnetic material. 
Iron can be alloyed or combined with other elements to form compounds that are metals, insulators~\cite{Badro1999}, superconductors~\cite{Hsu2008}, strong ferromagnetism~\cite{Kmmerle1977}, antiferromagnetism~\cite{Moruzzi1992}, thermal shielding at low and high temperatures and has good mechanical properties~\cite{Inoue2004}. 

Interest in magnetic materials is at an all-time high due to the end of Moore's law~\cite{Thompson2006} and other limitations that current electronic devices are facing. 
In that regard, magnetic materials are seen as an appealing solution leveraging the spin of the electron giving rise to the field of spintronics. 
Spintronic devices exploit both the electronic and spin degrees of freedom~\cite{uti2004,Han2023,Brataas2020}, offering promising logic and memory architectures with inherently low power consumption~\cite{Dieny2020,Vaz2024,BehinAein2010,Datta1990}. 
In addition, spintronic devices are widely used for memory applications~\cite{Hirohata2020,kerman2005,Bibes2008,Bhatti2017}. 
Iron and nickel are ubiquitous in steels, alloys, and functional magnetic devices used in the energy sector.
Thus, there is renewed effort to understand, search and characterize potential materials relevant for the mentioned applications~\cite{Baltz2018,Jungwirth2016,Meer2023,Gajek2007}. 
Apart from these advances in devices for classical computers, the field of quantum computing might also require magnetic compounds that present specific transport properties~\cite{Chumak2015,Han2019}, and a certain robustness to thermal noise to build and shield the quantum bits~\cite{DasSarma2005,Sau2010}.

Given that electrical and thermal conductivity in metals and semiconductors is strongly influenced by electron–phonon scattering, developing accurate first-principles methods to capture this mechanism is central to understanding and reducing energy losses in conductive materials.
Although electron-phonon interactions have been studied for decades, accurately describing them has proven to be a challenging task. 
The situation becomes even more complicated when one considers the effect that phonons can have on the local magnetic structure~\cite{Stupakiewicz2021,Nova2016,Ueda2023,Afanasiev2021,Davies2024}, where in some cases they can even induce a non-zero transient magnetization in nonmagnetic materials~\cite{Gurtubay2020,Basini2024}. 
As a result, it is common to rely on effective models that take input from experiments~\cite{ShirdelHavar2024,Streib2019,Cao2022}, or previously computed ab initio quantities~\cite{Dubrovin2024,FlrezGmez2022,Juraschek2020,Juraschek2021}.
Today, first-principles electron-phonon interactions are widely available thanks to density functional theory (DFT)~\cite{Hohenberg1964,Kohn1965} and density functional perturbation theory (DFPT)~\cite{Baroni1987,Gonze1997,Savrasov1992,Baroni2001}.
However, this approach is not free from challenges that can make actual calculations impractical. 
One of the main bottlenecks is the high computational cost required to sample the Brillouin zone in both electronic $\mathbf{k}$ and phononic $\mathbf{q}$ momentum space when integrating the different observables. 
In this regard, one of the most widespread solutions is to combine DFPT with maximally localized Wannier functions~\cite{Giustino2007,Marzari2012}, which enables interpolation at arbitrary momentum $\mathbf{k}$ and $\mathbf{q}$ of the electron-phonon matrix elements (EPME) $g_{mn\nu}(\mathbf{k},\mathbf{q})$, electronic eigenvalues $\varepsilon_{n\mathbf{k}}$ and eigenvectors $\psi_{n\mathbf{k}}$. 
Most implementations to date assume time-reversal symmetry. 
This assumption is only suited to the study of non-magnetic materials, hindering a global understanding of electron-phonon interactions. 
In fact, this is an important barrier to understanding the role of phonons in spintronics, caloritronics, or superconductivity and charge density waves in the presence of magnetism. 
More importantly, both electron and phonon dispersions can be affected by including the spin degree of freedom in the calculations while retaining the same exchange and correlation functional~\cite{Carrascolvarez2024}. 
For example, neglecting the spin of iron leads to unstable phonons~\cite{Ma2023}.

This limitation has been addressed using finite differences~\cite{Chaput2019,Engel2020,Carrascolvarez2024,Carrascolvarez2024b,Carrascolvarez2025,WangYan2024,Zhang2025,Poliukhin2025,Wang2026}
but is limited due to the computational cost of supercells. 
Recently, support for DFPT with non-collinear magnetism and ultrasoft pseudopotentials has been implemented in \textsc{Abinit}~\cite{Gonze2016,Gonze2020,Verstraete2013,Ma2023,Verstraete2025} and \textsc{Quantum ESPRESSO} (QE)~\cite{Giannozzi2009,Giannozzi2017,Urru2019}, as well as interpolation of the electron-phonon matrix elements with collinear magnetism in \textsc{Perturbo}~\cite{Zhou2021a,Zhou2021,Chang2025,Le2025}.
%
%
We here present a first-principles implementation for electron–phonon properties in magnetic systems, validated and applied to Fe and Ni. 
Our results reveal the essential role of the magnetic phase in resistivity, the microscopic origin of scattering in different ferromagnets, and the absence of phonon-mediated superconductivity in both materials. 
These insights provide a foundation for designing magnetic materials with tailored transport properties for energy applications.
In this work, we extend the capabilities of the electron phonon Wannier (\textsc{EPW}) code~\cite{Giustino2007,Ponce2016,Lee2023} to support collinear magnetism, unlocking the path to study electron-phonon related properties in magnetic materials with high accuracy and at reasonable computational cost.
To this end, we exploit recent developments in the \textsc{Abinit} code on spin-dependent electron-phonon interactions~\cite{Gonze2020} to validate our implementation. 
This choice is guided by the documented small discrepancy between \textsc{Abinit} and \textsc{Quantum ESPRESSO}~\cite{Ponce2014,Bosoni2023,Ponce2025}.
Although collinear magnetism serves as an approximation to its non-collinear counterpart when considering spin-orbit coupling (SOC), its computational efficiency is markedly superior, while retaining a good degree of accuracy.

We study the carrier transport properties of ferromagnetic (FM) body-centered cubic (BCC) Fe and face-centered cubic (FCC) Ni when electron phonon interactions are accounted. 
Our findings indicate that the omission of magnetism in Fe leads to the emergence of imaginary (soft) phonon modes, confirming previous works. 
As sometimes done in the literature~\cite{Mounet2018,Sohier2020,Zhang2025,Goudreault2025}, excluding imaginary phonon modes overestimated resistivity due to the large contribution of modes on the onset of being imaginary.
In contrast, Ni is a weaker ferromagnet with similar ferromagnetic and non-magnetic phonon band dispersions, but with a very different change of resistivity with temperature.
Overall, the inclusion of magnetism reduces resistivity by a factor of three and seven for Fe and Ni, respectively. 
Finally, we compute the phonon-driven superconducting transition temperature ($T_{\rm c}$) in Fe and Ni and find vanishingly small values, which shows that neither material would be superconducting, even if their Cooper pairs could survive in the FM magnetization field. 

This work opens the gate towards accounting for the role of phonons in magnetic systems and possible superconducting states in magnetic materials (such as iron arsenides and possibly cuprates), which are highly relevant for new energy-efficient technologies and energy loss in conducting materials.
This work is organized as follows: we first present the theory in Sect.~\ref{sec:theory}, details about our computational framework in Sect.~\ref{sec:method}, the interpolation of electron, phonon and electron-phonon quantities in Sect.~\ref{sec:elph}, the calculation of  electron-phonon properties in Sect.~\ref{sec:properties}, and the final remarks and conclusions in Sect.~\ref{sec:conclusions}.

\section{Theory}\label{sec:theory}

We start with the definition of the EPME in the phonon mode representation~\cite{Giustino2017}  
\begin{multline}
    g_{mn\nu}(\mathbf{k},\mathbf{q}) = \sum_{\kappa\alpha p}\sqrt{\frac{\hbar}{2M_\kappa\omega_{\mathbf{q}\nu}}}e_{\kappa\alpha\nu\mathbf{q}} \\
    \times \mel{\psi_{m\mathbf{k+q}}}{\partial_{\kappa\alpha,\mathbf{q}}V}{\psi_{n\mathbf{k}}}, 
\end{multline}
where $\omega_{\mathbf{q}\nu}$ is the frequency of the phonon mode $\nu$ with quasi-momentum $\mathbf{q}$, $e_{\kappa\alpha\nu\mathbf{q}}$ the associated eigenvector, $M_\kappa$ is the mass of the $\kappa$-th atom in the unit cell, and $\partial_{\kappa\alpha,\mathbf{q}}V$ is the variation of the effective potential with respect to the perturbation involving atom $\kappa$ displaced in Cartesian direction $\alpha$. 
This definition has to be updated when considering non-collinear magnetism as
\begin{multline}
    g^{\sigma'\sigma}_{mn\nu}(\mathbf{k},\mathbf{q})  = \sum_{\kappa\alpha p} \sqrt{\frac{\hbar}{2M_{\kappa}\omega_{\mathbf{q}\nu}}} e_{\kappa\alpha\nu\mathbf{q}}\\
    \times \mel{\psi^{\sigma'}_{m\mathbf{k+q}}}{\partial_{\kappa\alpha,\mathbf{q}}V^{\sigma'\sigma}}{\psi^{\sigma}_{n\mathbf{k}}} ,
\end{multline}
where $\sigma$ and $\sigma'$ are the spinor indices. 
The matrix element $g^{\sigma'\sigma}_{mn\nu}(\mathbf{k},\mathbf{q})$ can be written in terms of the deformation potential  $D^{\sigma'\sigma}_{mn\nu}(\mathbf{k},\mathbf{q})$ as~\cite{Ponce2021}
\begin{equation}
   \Big|g^{\sigma'\sigma}_{mn\nu}(\mathbf{k},\mathbf{q}) \Big| = \sqrt{\frac{\hbar}{m_\mathrm{u}\omega_{\mathbf{q}\nu}}} D^{\sigma'\sigma}_{mn\nu}(\mathbf{k},\mathbf{q}),
\end{equation}
with $m_\mathrm{u}$ being the atomic mass constant. 
$D^{\sigma'\sigma}_{mn\nu}(\mathbf{k},\mathbf{q})$ gives a measure of how much a phonon perturbation has changed the internal self-consistent potential felt by the electrons. 
When taking the absolute value of $|g^{\sigma'\sigma}_{mn\nu}(\mathbf{k},\mathbf{q})|$,
we take the root mean square average over any degenerate manifolds in $m$, $n$ and $\nu$. 
Note that the definitions of $D^{\sigma'\sigma}_{mn\nu}(\mathbf{k},\mathbf{q})$ and $g^{\sigma'\sigma}_{mn\nu}(\mathbf{k},\mathbf{q})$ are also valid for the case of nonmagnetic compounds with and without SOC. 
The main property exploited in the nonmagnetic case are the following extra symmetries in the spin indices $[g^{\sigma'\sigma}_{mn\nu}(\mathbf{k},\mathbf{q})]^* = (-1)^{\sigma+\sigma'} g^{\bar{\sigma}'\bar{\sigma}}_{mn\nu}(\mathbf{-k},\mathbf{-q})$, with $\bar{\sigma}$ being the opposite spin to $\sigma$. This means that if $\sigma = \sigma'$ then $[g^{\sigma'\sigma}_{mn\nu}(\mathbf{k},\mathbf{q})]^* = g^{\bar{\sigma}'\bar{\sigma}}_{mn\nu}(\mathbf{-k},\mathbf{-q})$, and if $\sigma\neq\sigma'$ then $[g^{\sigma'\sigma}_{mn\nu}(\mathbf{k},\mathbf{q})]^* = -g^{\bar{\sigma}'\bar{\sigma}}_{mn\nu}(\mathbf{-k},\mathbf{-q})$ (e.g. for heavy elements with spin orbit coupling but no magnetization).
These two results are a consequence of time-reversal symmetry, which is absent in magnetic compounds.
In practice, performing magnetic calculations with SOC, commonly known as non-collinear magnetic calculations, is computationally expensive (typically by a factor of $2^3=8$ for cubic scaling DFT) in systems with several atoms. 
The collinear approximation consists in separating the two spin channels into up ($\uparrow$) and down ($\downarrow$) without explicit interaction between them, thereby only doubling the cost of the calculations, regardless of the system size.  
This allows us to define a spin-resolved EPME: 
\begin{equation}
   g^{\sigma}_{mn\nu}(\mathbf{k},\mathbf{q}) \equiv \delta_{\sigma'\sigma} g^{\sigma'\sigma}_{mn\nu}(\mathbf{k},\mathbf{q})
\end{equation}
and it follows that the deformation potential becomes
\begin{equation}
    D^{\sigma}_{mn\nu}(\mathbf{k},\mathbf{q}) = \delta_{\sigma'\sigma}D^{\sigma'\sigma}_{mn\nu}(\mathbf{k},\mathbf{q}).
\end{equation}
In systems with many valence electrons, we are commonly interested in the deformation potential of several bands. 
For each spin channel, we can consider 
\begin{equation}\label{D-sum}
   D^\sigma_{\nu}(\mathbf{k},\mathbf{q}) = \sqrt{\sum_{mn \in \mathcal{M}}[D^\sigma_{mn\nu}(\mathbf{k},\mathbf{q})]^2} 
\end{equation}
as the deformation potential of interest, with the sum over the electronic bands $m,n$ restricted to the bands in the Wannier manifold $\mathcal{M}$.
We also extend the mode- and momentum-resolved electron-phonon coupling $\lambda_{\mathbf{q}\nu}$, related to the imaginary part of the phonon self energy $\Im\Pi_{\mathbf{q}\nu}$, which in the double delta approximation at the collinear level becomes~\cite{Moseni2024}
\begin{multline}\label{eq:lambda}
    \lambda_{\mathbf{q}\nu} \equiv \sum_\sigma\lambda_{\mathbf{q}\nu}^\sigma = \sum_{\sigma mn\mathbf{k}}\frac{w_{\mathbf{k}}}{N(\varepsilon_{\rm F})\omega_{\mathbf{q}\nu}}|g_{mn\nu}^{\sigma}(\mathbf{k},\mathbf{q})|^2\\
    \times \delta(\varepsilon_{\rm F}-\varepsilon^{\sigma}_{n\mathbf{k}})\delta(\varepsilon_{\rm F} - \varepsilon^{\sigma} _{m\mathbf{k+q}}) = \frac{1}{\pi N(\varepsilon_{\mathrm{F}})}\frac{\Im\Pi_{\mathbf{q}\nu} }{\omega^2_{\mathbf{q}\nu}},
\end{multline}
where $N(\varepsilon_{\textrm{F}})$ is the total density of states at the Fermi energy $\varepsilon_{\rm F}$ and $w_{\mathbf{k}}$ are the $\mathbf{k}$-point integration weights normalized to one. 
In the non-magnetic case, the sum over $\sigma$ is removed and $\sum_{\mathbf{k}} w_{\mathbf{k}} = 2$ due to time-reversal symmetry. 
We note that another definition exists for the spin-resolved electron-phonon coupling using the partial density of states at the Fermi level $N^{\sigma}(\varepsilon_{\rm F})$~\cite{Verstraete2013,Badrtdinov2023}. 
However, this convention will suffer from divergences in half- and semi-metals, where only one spin channel contributes to the total density of states. 
The present version should be used systematically for spin polarized cases, taking into account the factor of 2 between $N^{\sigma}$ and $N$ in the unpolarized case.

Another important quantity is the mode- and momentum-resolved phonon linewidth $\gamma_{\mathbf{q}\nu}$ 
\begin{multline}
    \gamma_{\mathbf{q}\nu} \equiv \sum_\sigma\gamma^\sigma_{\mathbf{q}\nu} = 2\pi\omega_{\mathbf{q}\nu}\sum_{\sigma\mathbf{k}mn}w_{\mathbf{k}}|g^\sigma_{mn\nu}(\mathbf{k},\mathbf{q})|^2 \\
    \times \delta(\varepsilon_{\rm F} - \varepsilon^\sigma_{n\mathbf{k}})\delta(\varepsilon_{\rm F} - \varepsilon^\sigma_{m\mathbf{k+q}}), 
\end{multline}
which can be used to compute the Eliashberg spectral function
\begin{equation}
    \alpha^2F(\omega) = \frac{1}{2\pi N(\varepsilon_{\rm F})}\sum_{\sigma\mathbf{q}\nu}w_{\mathbf{q}}\frac{\gamma^\sigma_{\mathbf{q}\nu}}{\omega_{\mathbf{q}\nu}}\delta(\omega-\omega_{\mathbf{q}\nu}),
\end{equation}
where $w_{\mathbf{q}}$ are the $\mathbf{q}$-point integration weights normalized to one.
The Eliashberg spectral function can be used to determine the contribution of specific phonon modes to the total electron-phonon coupling strength defined as
\begin{equation}
    \lambda \equiv \int \frac{d\omega}{\omega}\alpha^2F(\omega),
\end{equation}
This quantity provides a measure of the overall mass enhancement in a metallic system. 

Finally, we study resistive transport in magnetic metals.
The transport of charge in a magnetic material depends on the spin-resolved carrier scattering rate $1/\tau_{n\mathbf{k}}^\sigma$ defined as
\begin{multline}\label{eq:scattering}
        \frac{1}{\tau_{n\mathbf{k}}^\sigma} \equiv \sum_{\mathbf{q}m\nu}w_{\mathbf{q}}|g_{mn\nu}^{\sigma}(\mathbf{k},\mathbf{q})|^2 \\
        \times[(n_{\mathbf{q\nu}}+1-f_{m\mathbf{k+q}}^{\sigma})\delta(\varepsilon_{n\mathbf{k}}^{\sigma}-\varepsilon_{m\mathbf{k+q}}^{\sigma}-\hbar\omega_{\mathbf{q}\nu}) \\
        +(n_{\mathbf{q\nu}}+f_{m\mathbf{k+q}}^{\sigma})\delta(\varepsilon_{n\mathbf{k}}^{\sigma}-\varepsilon_{m\mathbf{k+q}}^{\sigma}+\hbar\omega_{\mathbf{q}\nu})].
\end{multline}
where $w_{\mathbf{q}}$ follow the same integration rules as $w_{\mathbf{k}}$, $f_{m\mathbf{k+q}}^{\sigma}$ is the Fermi-Dirac occupation for spin $\sigma$ and $n_{\mathbf{q\nu}}$ is the Bose-Einstein distribution function. 
As for previous quantities, the average scattering rate is additive in the collinear approximation to magnetism:
\begin{equation}\label{mean-tau}
    \langle{\tau}^{-1}\rangle \sim \sum_{\sigma}\sum_{n\mathbf{k}}w_\mathbf{k} \frac{1}{\tau_{n\mathbf{k}}^\sigma}.
\end{equation}

Previous studies have used the lowest-order variational approximation (LOVA) to study the resistivity of magnetic systems~\cite{Ma2023}, which is based on computing the average scattering rate. 
This is a strong approximation in anisotropic systems or systems with different Fermi sheets~\cite{Ponce2020}. 
Here, we solve the linearized Boltzmann transport equation (BTE) to calculate the conductivity~\cite{Ponce2018,Claes2025}
\begin{equation}
    \sigma_{\alpha\beta}^{\sigma} = \frac{-e}{V^{\rm uc}}\sum_{n\mathbf{k}}w_\mathbf{k}v^{\sigma}_{n\mathbf{k}\alpha} \partial_{E_\beta} f_{n\mathbf{k}}^\sigma,
\end{equation}
where $V^{\rm uc}$ is the volume of the unit cell, $\partial_{E_\beta} f_{n\mathbf{k}}^\sigma$ is the perturbed occupation factor due to an applied electric field $E_\beta$ in the direction $\beta$, (see, Eq.~\eqref{eq:ibte} below) $v_{n\mathbf{k}\alpha}^{\sigma}$ is the $\alpha$ component of the spin-resolved electronic band velocity, and $\sigma_{\alpha\beta}^{\sigma}$ is the conductivity tensor for spin index $\sigma$. 
The corresponding resistivity is 
\begin{equation}\label{eq:rhoBTE}
    \rho_{\alpha\beta}^{\rm BTE} \equiv (\sigma_{\alpha\beta}^\uparrow+\sigma_{\alpha\beta}^\downarrow)^{-1} = \left[(\rho_{\alpha\beta}^\uparrow)^{-1}+(\rho_{\alpha\beta}^\downarrow)^{-1}\right]^{-1},
\end{equation}
which is the usual law of addition for resistivities in parallel, and in this case $\rho_{\alpha\beta}^\sigma$ is the resistivity experienced by the electrons in each spin channel. 
Interestingly, when using the Ziman or LOVA formula, the resistivity becomes additive instead $\rho_{\alpha\beta}^\uparrow+\rho_{\alpha\beta}^\downarrow$,
questioning the validity of Eq.~(8) of Ref.~\cite{Ma2023}, see Supplementary Material S1 for more information.
When comparing to polycrystalline samples, we use 
\begin{equation}
    \rho^{\rm BTE} = \frac{\rm{Tr}[\rho_{\alpha\beta}^{\rm BTE}]}{3}.
\end{equation}
The out-of-equilibrium occupations $\partial_{E_\beta}f_{n\mathbf{k}}^\sigma$~\cite{Ponce2020} in the collinear magnetic case are obtained by solving the BTE
%
\begin{multline}\label{eq:ibte}
    \partial_{E_\beta} f_{n\mathbf{k}}^\sigma = ev_{n\mathbf{k}\beta}^{\sigma}
    \frac{\partial f^{\sigma}_{n\mathbf{k}}}{\partial \varepsilon^\sigma_{n\mathbf{k}}}
    \tau_{n\mathbf{k}}^\sigma+\frac{2\pi\tau_{n\mathbf{k}}^\sigma}{\hbar}\sum_{\mathbf{q}m\nu}w_{\mathbf{q}}|g_{mn\nu}^{\sigma}(\mathbf{k},\mathbf{q})|^2 \\
    \times[(n_{\mathbf{q\nu}}+1-f_{n\mathbf{k}}^{\sigma})\delta(\varepsilon_{n\mathbf{k}}^{\sigma}-\varepsilon_{m\mathbf{k+q}}^{\sigma}-\hbar\omega_{\mathbf{q}\nu}) \\\!+\!(n_{\mathbf{q\nu}}\!+\!f_{n\mathbf{k}}^{\sigma})\delta(\varepsilon_{n\mathbf{k}}^{\sigma}\!-\!\varepsilon_{m\mathbf{k+q}}^{\sigma}\!+\!\hbar\omega_{\mathbf{q}\nu})]\partial_{E_\beta} f_{m\mathbf{k+q}}^\sigma.
\end{multline}
Here $\partial f^\sigma_{n\mathbf{k}} / \partial\varepsilon^\sigma_{n\mathbf{k}}$ is the derivative of the Fermi-Dirac distribution. 
The equation is solved through an iterative procedure and if we neglect the second term in Eq.~\eqref{eq:ibte}, we obtain the self-energy relaxation time approximation (SERTA)~\cite{Ponce2018}, which in the case of metallic systems is usually quite accurate.

\section{Methods}\label{sec:method}

DFT calculations are performed using \textsc{QE}~\cite{Giannozzi2009,Giannozzi2017} version 7.4 and \textsc{Abinit} version 10.2.5~\cite{Gonze2002,Gonze2016,Gonze2020,Verstraete2025}. 
Scalar-relativistic norm-conserving pseudopotentials~\cite{Hamann2013} from the standard accuracy table of the \textsc{PseudoDojo} project~\cite{vanSetten2018} are used.
We use the local density approximation (LDA)~\cite{Perdew1981} and the generalized gradient approximation (GGA) with the Perdew Burke Ernzerhof (PBE) functional~\cite{Perdew1996}.
The interpolated phonon dispersions and dynamical matrices are obtained with DFPT~\cite{Baroni1987,Savrasov1992,Gonze1997,Baroni2001}. 
The maximally localized Wannier functions are obtained using the \textsc{wannier90} package version 3.1~\cite{Pizzi2020}. 
Direct evaluation of spin-resolved EPME is obtained by modifying \textsc{QE} to print the values.
We provide the modified code on the Materials Cloud Archive~\cite{MCA2025} along with all data to reproduce the results presented here.
The interpolation of electronic band structures, phonon band structures, and electron phonon matrix elements, as well as the calculation of transport properties and electron-phonon coupling matrix elements are performed by modifying the \textsc{EPW} package version 5.9~\cite{Lee2023}. 
This feature will be released in a future \textsc{EPW} version and we also provide the code on the Materials Cloud Archive~\cite{MCA2025}.

The cutoff energy for the plane wave expansion, $\bf k$-meshes for the electrons, and the $\bf q$-meshes for the phonons are selected so that an arbitrary increase in those parameters produces a total energy change of less than 1~meV per unit cell.
This corresponds to 100~Ry for both Fe and Ni, as well as a $\bf k$-point grid of 24$^3$ and a $\bf q$-point grid of 8$^3$ for both Ni and Fe.
Structural relaxations are performed with a convergence criterion of $\Delta E < 10^{-12}$~Ry, and with a pressure of less than 1~bar. 
In the case of phonons, the convergence criteria in order to obtain accurate force constants was \texttt{tr2\_ph} $= 10^{-20}$ for \textsc{QE} while for \textsc{Abinit} the convergence criteria is on the energy $\Delta E < 10^{-20}$~Ha.
We converge the resistivity below 1\% above 150~K using \textsc{EPW}, and validate the results with an in-house developed version of the \textsc{ElectronPhonon.jl} package~\cite{Lihm2024, EPjl}, which relies on the real-space EPME computed with the \textsc{EPW} code~\cite{Ponce2016,Lee2023} that we extend to the collinear magnetic case for the purpose of this work. 
Excellent agreement is obtained between the two codes, see Supplementary Material S2~\cite{MCA2025} for more information.  

\section{Electron-phonon coupling with magnetism}\label{sec:elph}

In this work, we study BCC Fe and FCC Ni, both in their ferromagnetic configuration using the collinear approximation.
We also consider non-magnetic Pb and verify that the LSDA implementation yields the same results as the LDA case as described in Supplementary Material S3~\cite{MCA2025}.

\subsection{Verification of the direct evaluation of the electron-phonon coupling}

To verify our implementation, we compare the \textsc{QE} results with those obtained with \textsc{Abinit}. 
Although both codes match well in non-magnetic compounds~\cite{Ponce2014,Bosoni2023,Ponce2025}, no validation study had yet been performed in the magnetic case.
To this end, we compare basic quantities including total energy, total magnetization and phonon frequencies while keeping the computational parameters as close as possible.
We find an excellent level of agreement between the two codes that is similar to the non-magnetic case, see Supplementary Material S4~\cite{MCA2025}. 

\begin{figure}[t]
    \centering
    \includegraphics[width=0.95\linewidth]{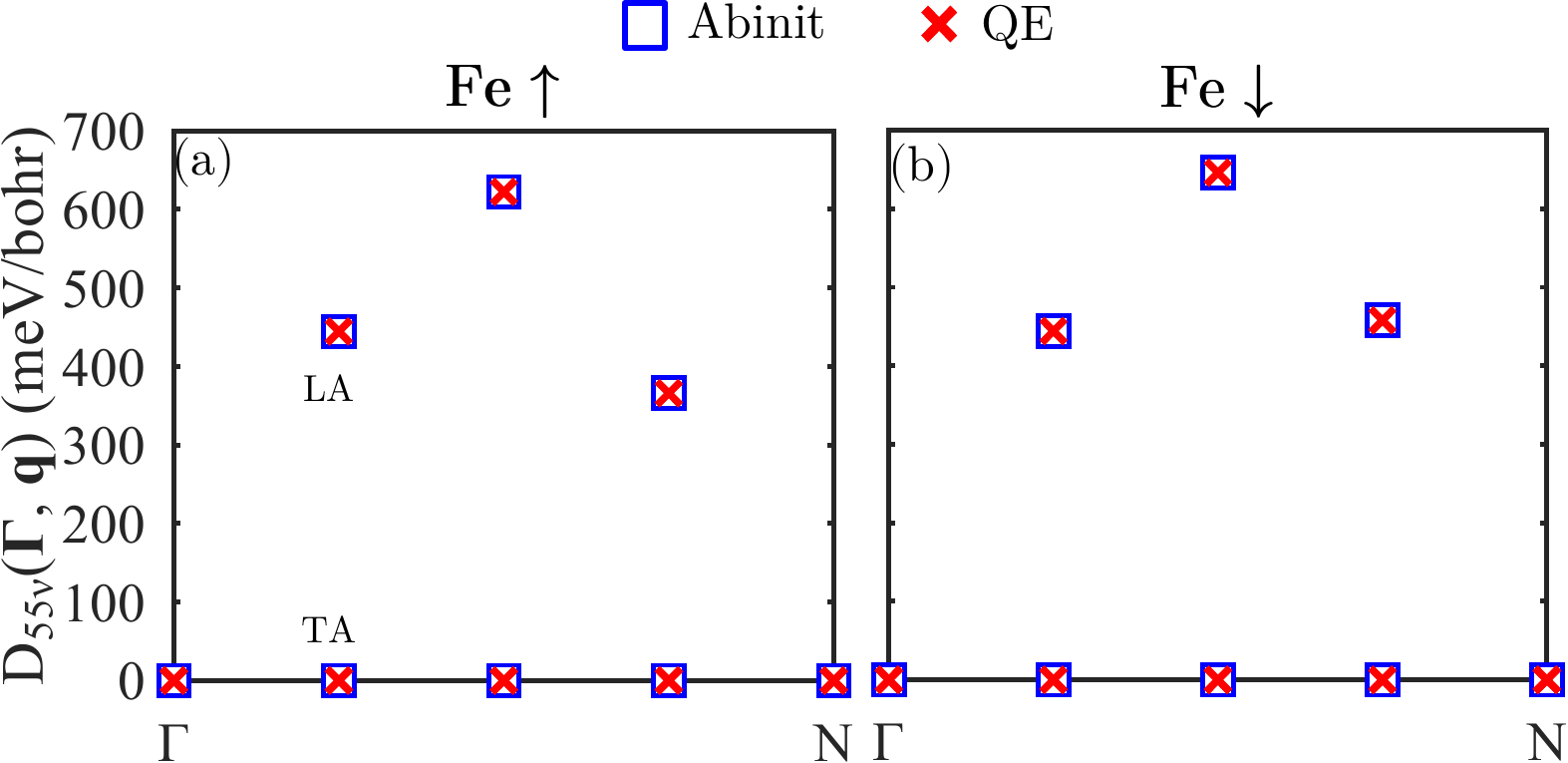}
    \caption{\label{g-QEvsAbinit}
    Direct evaluation of ferromagnetic iron deformation potential $D^{\sigma\sigma}_{m=5,n=5,\nu}(\mathbf{k}=\Gamma,\mathbf{q})$ for (a) spin up and (b) spin down at $\mathbf{k} = \Gamma$ for $m = n = 5$ along a $\mathbf{q}$ momentum line from $\Gamma$ to $\mathrm{N}$ for all phonon modes $\nu$ (LA and TA). 
    Red crosses are computed with QE while blue squares are computed with \textsc{Abinit}. 
    The LDA functional is used.
}
\end{figure}
 
We compare the absolute value of the deformation potentials between the two codes. 
To achieve this, we note that \textsc{QE} performs an average over all degenerate states in electronic and phonon band indices.
We therefore applied the same average to the \textsc{Abinit} results and present a comparison for ferromagnetic Fe in Fig.~\ref{g-QEvsAbinit} between the two codes for $\mathbf{k}=\boldsymbol{\Gamma}$ and bands $m = n = 5$.
We find a good agreement between the two codes with a maximum error of less than 0.02~meV/bohr. 
We provide the same comparison for Ni in Supplemental Material S5~\cite{MCA2025} where we find an error below 0.01~meV/bohr.

\subsection{Verification of the electron phonon interpolation}

Since we obtained reliable EPME by direct evaluation in \textsc{QE}, we use the \textsc{EPW} package to interpolate them into finer $\mathbf{q}$ and $\mathbf{k}$ grids. 
We extend the \textsc{EPW} package to be compatible with collinear magnetism and introduce two new input variables to control the selection of spin channel.  
\begin{figure}[t]
    \centering
    \includegraphics[width=0.99\linewidth]{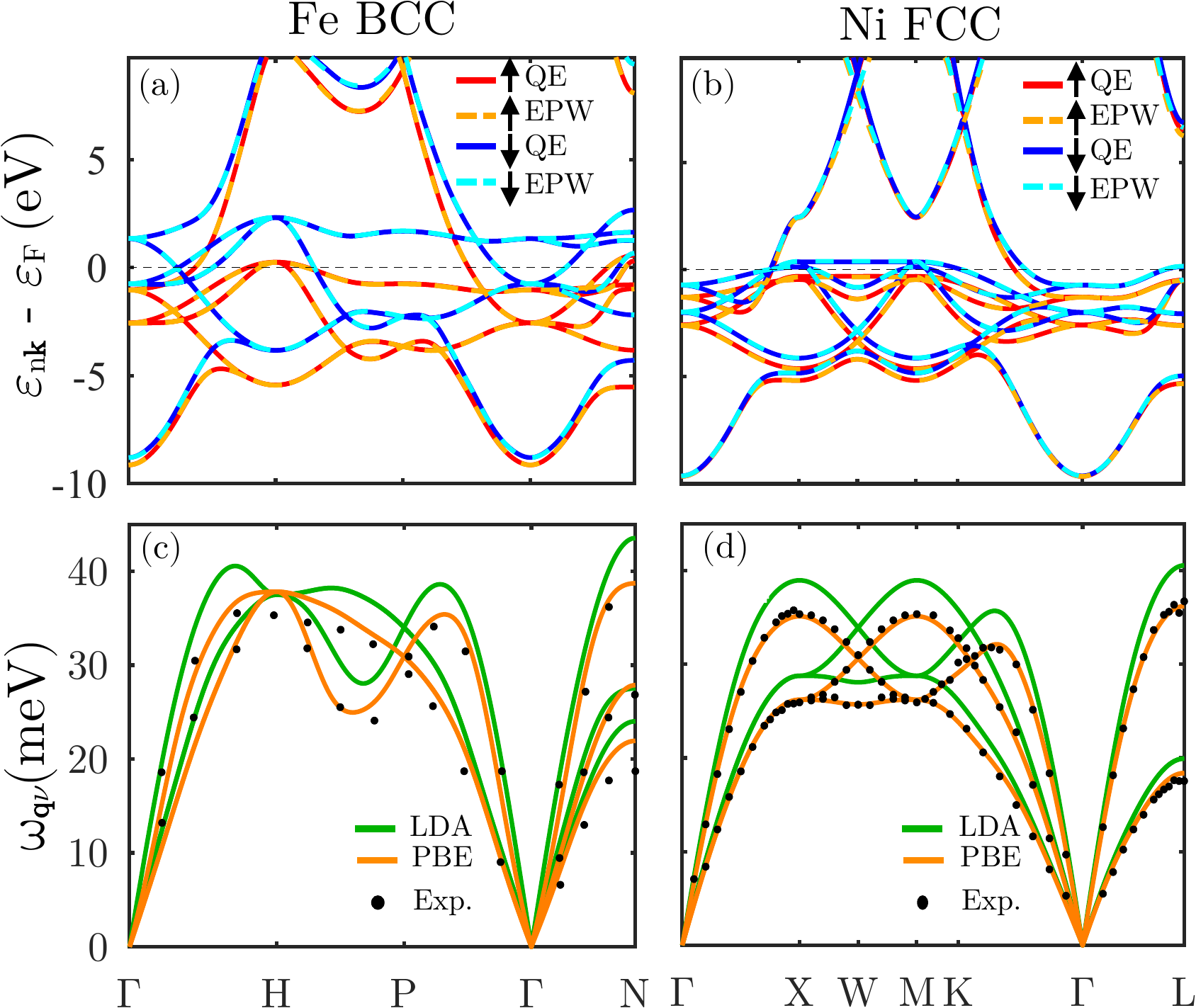}
    \caption{\label{band-interp}  Electronic band structure for ferromagnetic (a) Fe and (b) Ni. Solid (dashed) lines correspond to \textsc{QE} (\textsc{EPW}) calculations.
    Red/orange (blue/cyan) lines correspond to spin $\uparrow$ ($\downarrow$) or majority (minority) channel. 
    Phonon band structure for (c) Fe and (d) Ni with LDA (green) and PBE (orange) functionals where the black dots are experimental values from Refs.~\cite{Birgeneau1964,Brockhouse1967}.
    }
\end{figure}

\begin{figure}[t]
    \centering
    \includegraphics[width=0.99\linewidth]{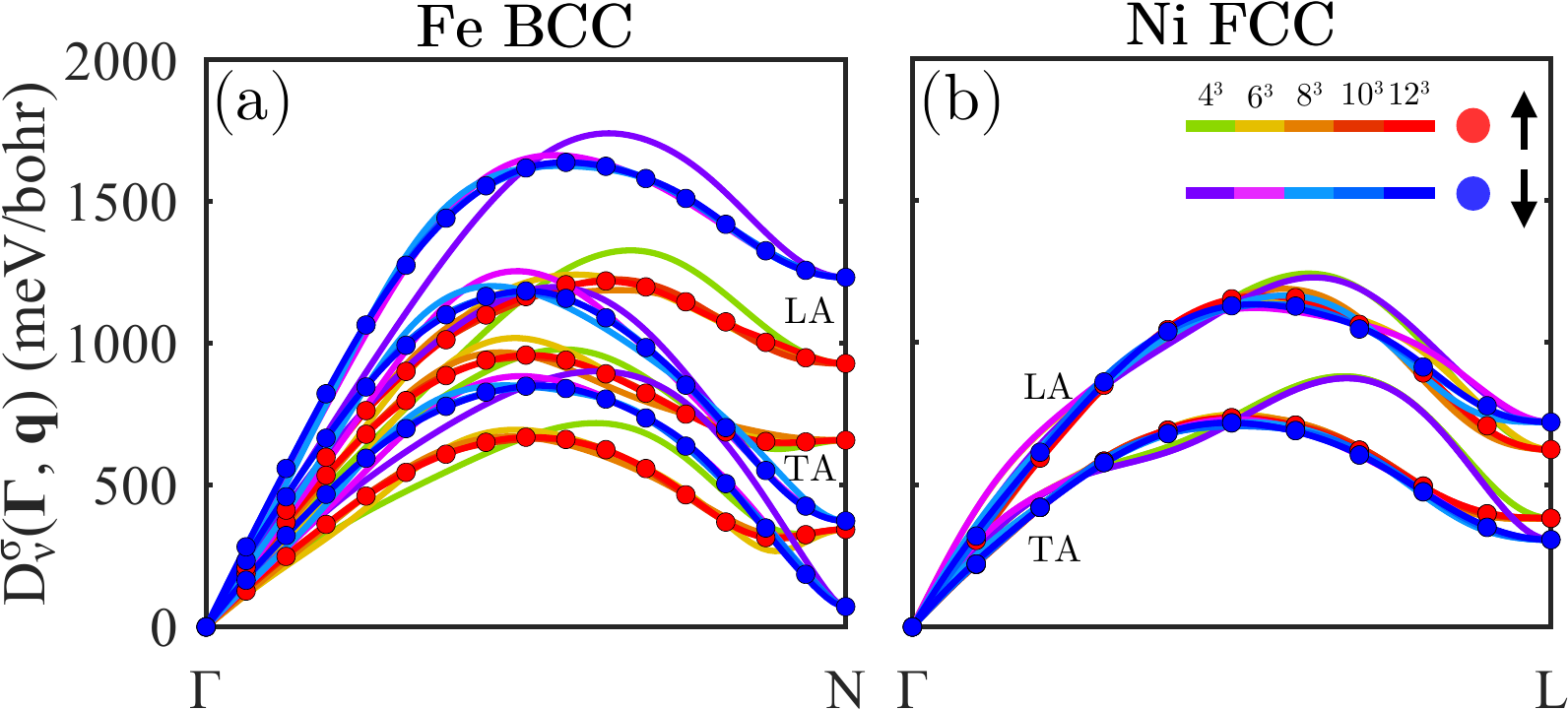}
    \caption{\label{g-interp}
    Spin resolved deformation potential for the three phonon branches $\nu$ (LA and TA)  at $\mathbf{k} = \Gamma$ and as a function of wave vector $\mathbf{q}$.
    The color scheme is presented as an inset in the figure $\uparrow$ (red/orange/yellow/green) and $\downarrow$ (blue/pink/purple). 
    The solid dots are direct evaluation of the deformation potential with QE. 
    The lines are the \textsc{EPW} interpolated deformation potential from $4^3$, $6^3$, $8^3$, $10^3$ and $12^3$ $\mathbf{k/q}$ homogeneous coarse grids with the colors provided in the inset of the figure.
    The electronic bands $m$ and $n$ are summed using Eq.~\eqref{D-sum} from second to sixth lowest electronic bands shown in Fig.~\ref{band-interp}(a) and \ref{band-interp}(b) for (a) Fe and (b) Ni. 
    }
\end{figure}
We first compare in Fig.~\ref{band-interp}(a) and \ref{band-interp}(b) the direct and Wannier-interpolated electronic band structures of ferromagnetic Fe and Ni respectively, obtaining an energy mismatch of less than 0.1~meV for all the displayed $\mathbf{k}$-points and bands. 
This shows that both spin channels are correctly reproduced by the interpolation procedure.
We then compute in Fig.~\ref{band-interp}(c) and \ref{band-interp}(d) the phonon dispersion of ferromagnetic Fe and Ni respectively using LDA and PBE.
Overall, LDA overestimates by about 10\% the phonon frequency with respect to neutron scattering experiments~\cite{Birgeneau1964,Brockhouse1967}. 
Using the PBE functional yields a much better agreement, in line with a previous study~\cite{DalCorso2000}.
Moreover, we validate the interpolated electron-phonon deformation potential by comparing to direct calculations in Fig.~\ref{g-interp}, for electronic bands crossing the Fermi level and at the electronic zone center. 
We show the two spin channels for Fe and Ni, and find that convergence is achieved for 12$\times$12$\times$12 $\mathbf{k}/\mathbf{q}$ coarse grids in both cases. 
For such grids, we find excellent agreement between interpolated and independently calculated reference direct calculations. 
In addition, we also compare the \textsc{EPW} interpolation with the \textsc{Abinit} interpolation finding similar results; see Supplementary Material S5~\cite{MCA2025}. 
Finally, we verify that the Hamiltonian, dynamical matrices, and EPME all decay rapidly in real space; see Supplementary Material S6~\cite{MCA2025}.

\section{Physical properties and electron-phonon quantities}\label{sec:properties}

Having validated our spin-resolved Wannier-interpolation scheme in magnetic materials, we compute physical observables linked with electron-phonon interactions.  
We start by computing the full electron-phonon coupling strength $\lambda$, and the corresponding spin-resolved $\lambda^\sigma$ 
using Eq.~\eqref{eq:lambda}
\begin{equation}\label{lambda-spin}
    \lambda^\sigma = \sum_{\mathbf{q}\nu} w_{\mathbf{q}}\lambda_{\mathbf{q}\nu}^{\sigma}
\end{equation}
and we report the results for ferromagnetic Fe and Ni in Table~\ref{tab-lambda}.
\begin{table}[t]
    \centering
    \begin{tabular}{c c c c c}
    \hline\hline
         Compound & Magnetism & $\lambda$  & $\lambda^\uparrow$ & $\lambda^\downarrow$ \\ 
         \hline
        Fe (BCC) & FM & 0.2961 & 0.2180 & 0.0781 \\
        Ni (FCC) & FM & 0.2767 & 0.0040 & 0.2727 \\ 
 \hline\hline
    \end{tabular}
    \caption{Total and spin-resolved electron phonon coupling strength $\lambda$ for ferromagnetic (FM) Fe and Ni.
    }
    \label{tab-lambda}
\end{table}
\begin{figure}[b]
    \centering
    \includegraphics[width=0.99\linewidth]{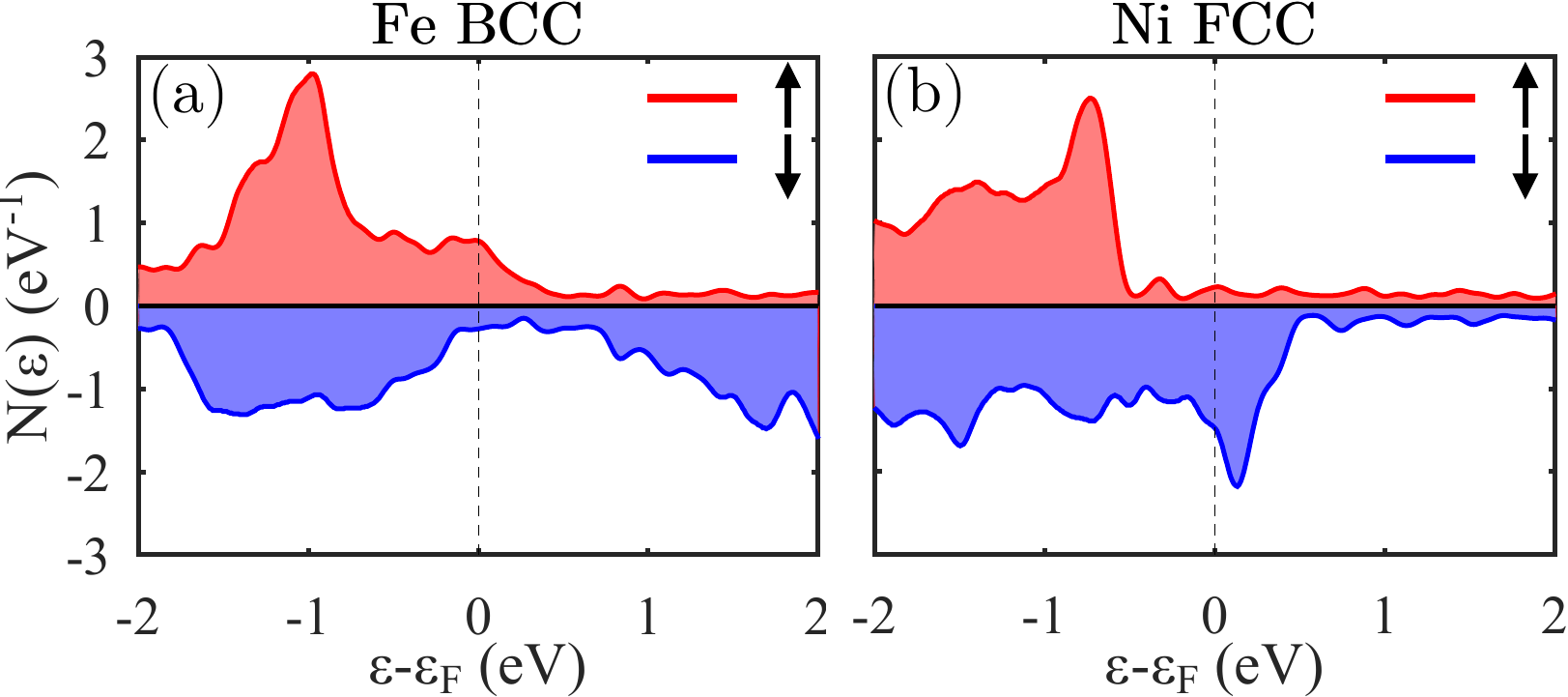}
    \caption{Spin resolved density of states for ferromagnetic (a) Fe and (b) Ni.
    Positive and negative density of states representing up (red) and down (blue) spin channels.}
    \label{DOS}
\end{figure}
Interestingly, we find that the calculated spin-resolved electron phonon coupling strength shows large differences between the two spin channels in both Fe and Ni. 
Although this behavior is not entirely surprising for Fe given the clear difference in deformation potential between the two spin channels, the same cannot be said in the case of Ni, whose splitting in deformation potential is much weaker, likely due to its smaller exchange splitting.
Additionally, we note that the main contribution to the electron phonon coupling strength in the case of Fe comes from the spin up (majority) channel, while in the case of Ni we find the reverse situation. 
In order to understand this behavior, we inspect the definition of the spin-resolved electron-phonon coupling strength Eqs.~\eqref{eq:lambda} and \eqref{lambda-spin} and we notice two main sources for this discrepancy. 
The first one is the deformation potential. 
The second is the electronic nesting for the given spin channel provided by the product of the two delta functions. 
This quantity is related to the density of states at the Fermi level. 
As seen in Fig.~\ref{g-interp}, the deformation potential is very different between the two spin channels for both Fe and Ni.
In the case of Fe the difference in the deformation potential is significant and could explain the difference.
However, $\lambda^\downarrow$ is smaller than $\lambda^\uparrow$, while the trend in the deformation potential is opposite, see Fig.~\ref{g-interp}(a). 
Consequently, the deformation potential cannot be the source of the difference between $\lambda^\uparrow$ and $\lambda^\downarrow$.
In the case of Ni, both spin channels have similar values in the deformation potential, while the electron phonon coupling strength $\lambda$ is completely dominated by the spin down channel $\lambda \approx \lambda^\downarrow$. 
Therefore, the deformation potential is not the main driver for such differences in either of the two considered cases.

Inspecting the density of states close to the Fermi level for each compound in Fig.~\ref{DOS}, we note that for Fe the spin up channel (majority) is 2.3 times larger than the spin down channel (minority), see Fig.~\ref{DOS}(a), while the opposite behavior is found for Ni where the spin down (minority) channel is 7.6 times larger, see Fig.~\ref{DOS}(b). 
This is a well-known effect linked to the $d$-band filling.
Thus, the differences in the spin-resolved electron-phonon coupling strength $\lambda^\sigma$ between the two spin channels for Fe and Ni are mainly related to Fermi-level nesting and the density of states at the Fermi level.
Our results for the electron-phonon coupling strength of Fe compare well with those obtained in Ref.~\cite{Moseni2024}.
\begin{figure}[t]
    \centering
    \includegraphics[width=0.99\linewidth]{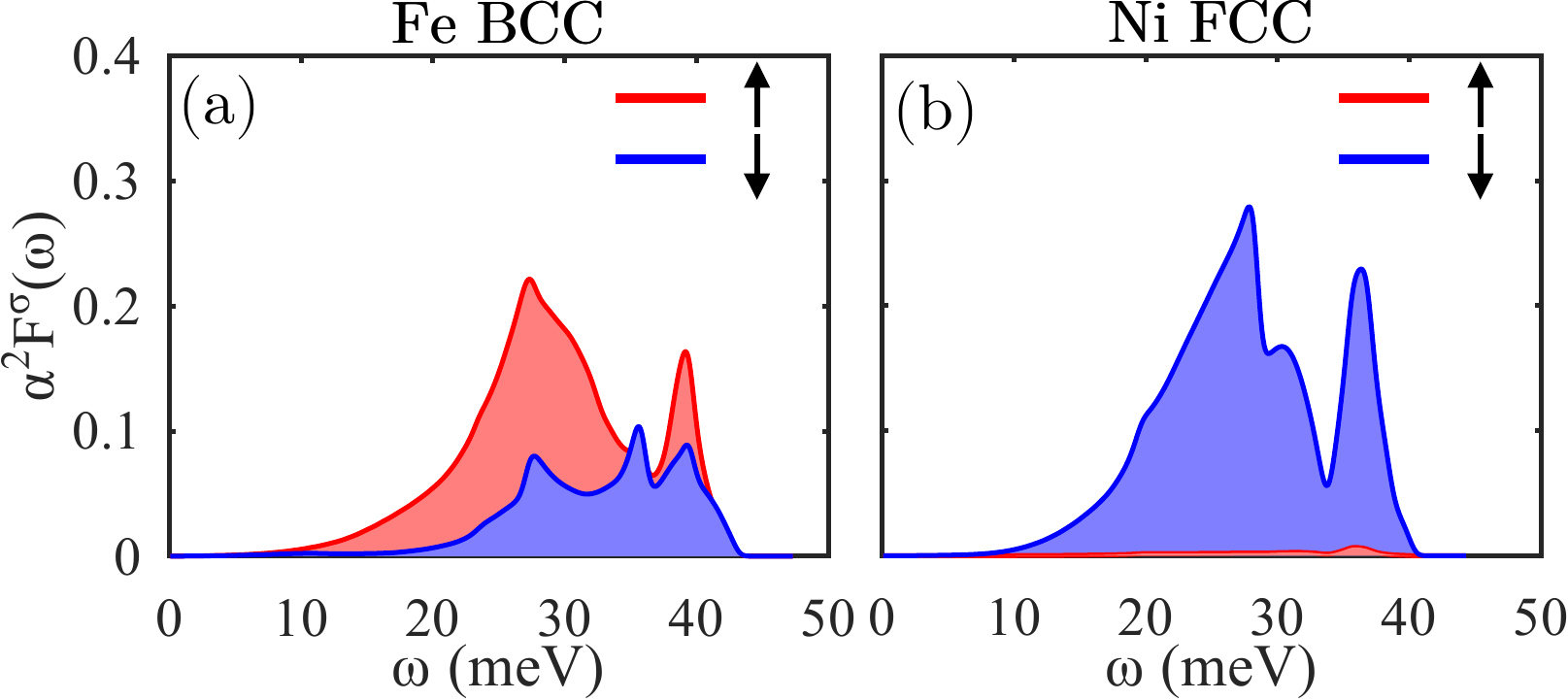}
    \caption{\label{fig:a2F}
    Eliashberg spectral function for ferromagnetic  (a) Fe and (b) Ni. 
    Red lines corresponds to the spin up (majority) and blue line corresponds to spin down (minority).
    }
\end{figure}

We also compute the Eliashberg spectral functions and show the result in Fig.~\ref{fig:a2F}. 
We see in Fig.~\ref{fig:a2F}(a) that the Fe spin majority channel is the dominant contribution for most of the energy range.  
In contrast, for Ni the spin minority dominates throughout and the spin majority is negligible, see Fig. ~\ref{fig:a2F}(b).
This important difference is due to the density of states at the Fermi level in Ni, see Fig.~\ref{DOS}. 
Recently, some authors have explored phonon-mediated superconductivity in magnetic systems by means of the mass enhancement $\lambda$ and the Eliashberg spectral function~\cite{Zhang2025}. 
Although the validity of using the Allen-Dyne formula~\cite{Allen1975} in magnetic materials has not yet been demonstrated, 
we report superconducting results for the non-magnetic and magnetic cases in Supplementary Material S7~\cite{MCA2025}.
In both cases and for both materials, we find a vanishing superconducting temperature in agreement with experimental observation.

\subsection{Resistivity of magnetic metals}

We now focus on the resistivity of metals and solve the BTE for both spin channels in ferromagnetic Fe and Ni. 
We note that when converging the resistivity and in general electron-phonon related quantities, it is possible to use different $\mathbf{k}/\mathbf{q}$ coarse and fine grids for the two spin channels since there is no guarantee that both spin channels would behave equivalently. 
Nevertheless, in our case, we find that convergence is achieved with the same settings. 
A good compromise between accuracy and computational cost is obtained with a $8^3$ $\mathbf{k/q}$ coarse grid; see Supplementary Material S8~\cite{MCA2025} for further discussion.

\begin{figure}[t]
    \centering
    \includegraphics[width=0.99\linewidth]{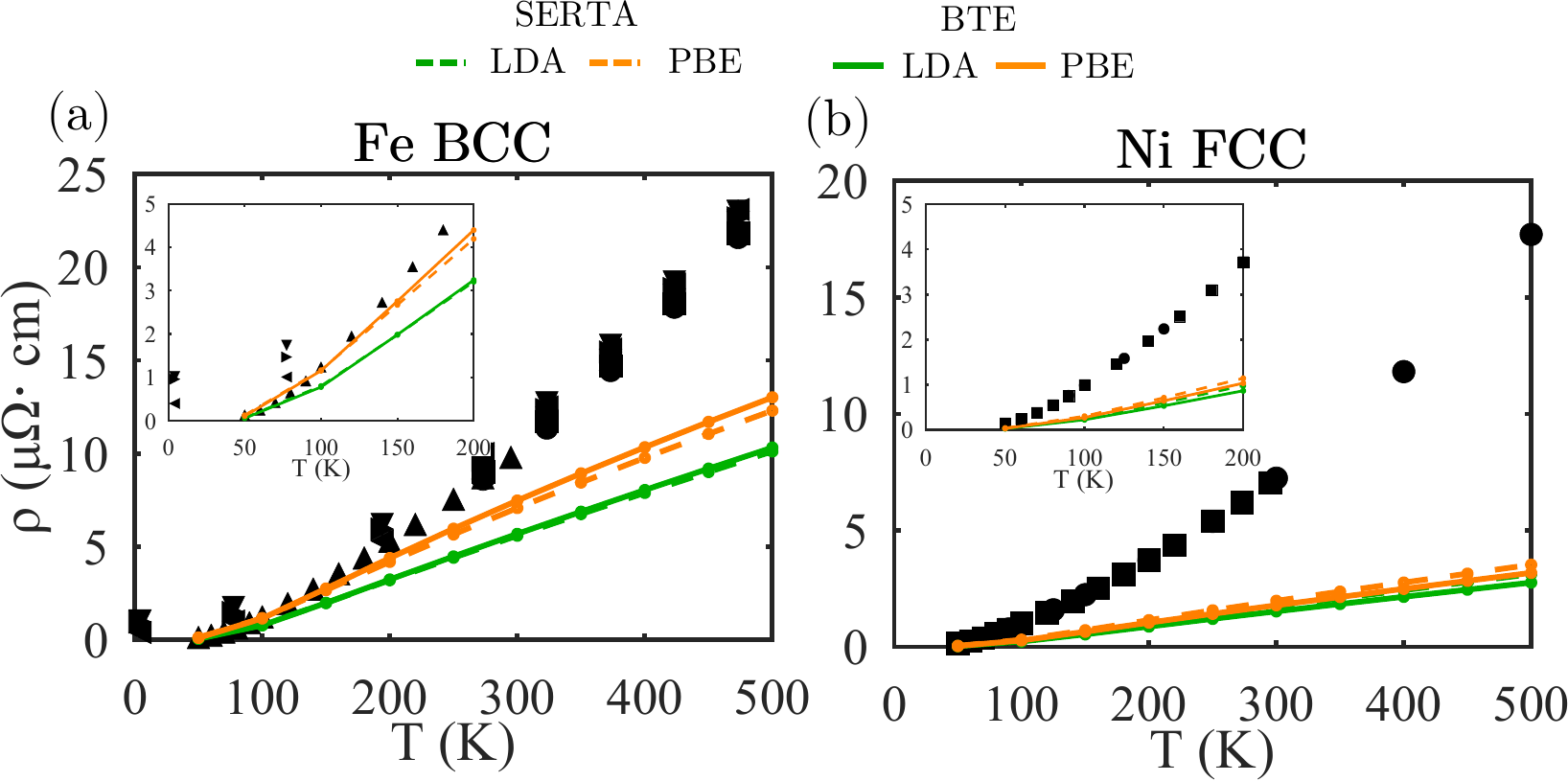}
    \caption{Electrical resistivity as a function of temperature for (a) Fe and (b) Ni, with an inset at low temperature 0-200~K.  
    Black symbols are obtained from Refs.~\cite{1959,Laubitz1976,Fulkerson1966}. 
    Lines are computed resistivities using SERTA (dashed) and full BTE (solid) with LDA (green) and PBE (orange).
    }
    \label{lda-res}
\end{figure}

The computed resistivity of Fe is presented in Fig.~\ref{lda-res}(a) and shows that the BTE with electron-phonon scattering explains only partially the resistivity and tends to underestimate it. 
We find that PBE results are slightly closer to experiment due to the improved phonon dispersion, see Fig.~\ref{band-interp}(c),  and a small decrease in Fermi velocity; see Supplementary material S9~\cite{MCA2025}.
We also find that the SERTA is a good approximation to the iterative BTE solutions, which is typical for non-magnetic metals~\cite{Lihm2025}. 
Our results are qualitatively similar to a previous study using the Ziman formula~\cite{Ma2023} where quantitative agreement with the experimental results is obtained at low temperatures.
However, we notice that the electron-phonon contribution to resistivity does seem to dominate at low temperature, see inset in Fig~\ref{lda-res}(a), and at temperatures up to 300~K our calculations suggest that electron-phonon scattering can account for at least 75\% of the experimental resistivity when using the PBE functional. 
For higher temperatures, the observed underestimation indicates that electron-phonon scattering is not the sole mechanism contributing to the resistivity.

We continue our analysis by investigating the case of ferromagnetic Ni. 
As seen in Fig.~\ref{lda-res}(b), the computed resistivity for Ni is also underestimated in the entire temperature range compared to experiment. 
However, above 350~K the resistivity originating from electron-phonon scattering is almost one order of magnitude less than the experimental resistivity. 
In addition, the agreement at low temperatures is worse than in Fe since there is an important underestimation of the resistivity already in the 50-100~K temperature range, see the inset in Fig.~\ref{lda-res}(b). 
Interestingly, we note that using the PBE functional does not improve significantly in spite of the remarkably better agreement of the phonon dispersion with experiments, see Fig.\ref{band-interp}(d). 
This hints at an important distinction with respect to Fe.\\ 

In ferromagnetic compounds, contributions to the resistivity originating from magnons or spin fluctuations are expected to be significant~\cite{Raquet2002,Watzman2016}. 
One of the reasons why magnons may explain the missing resistivity of Ni and Fe above 300~K is due to their energy scale which is around 8 to 9 times bigger than the phonons~\cite{Haliov1998}. 
The magnon population scales as $(T/T_{\rm Curie})^2$~\cite{Watzman2016}, and will be smaller in Fe than in Ni for a given temperature due to the lower Curie temperature ($T_{\rm Curie}$) of 627~K ~\cite{Laubitz1976} in Ni, compared to 1043~K in Fe~\cite{Fulkerson1966}. 
Consequently, while at 300~K the phonon contribution to the resistivity would show a linear ``high temperature behavior" , the magnon contribution would still be within its ``low temperature regime'' with a $T^2$ behavior.
Although this explains the increase in the importance of magnons at temperatures beyond 200-300~K, it does not explain the marked difference between Fe and Ni.

In fact, the magnon population is crucially dependent on the density of states at the Fermi level of the \emph{opposite} spin channel~\cite{Muller2019}.
The large density of states of the spin down (minority) channel in Ni (see Fig. \ref{DOS}(b)) above the Fermi level allows for a large magnon creation by converting spin up (majority) electrons into spin down (minority) electrons. 
In the case of Fe, magnon creation is suppressed by the small accessible phase space of minority electrons~\cite{Muller2019}. 
In addition to the difference in magnon creation, our calculations show that the spin up electrons in Ni experience far less scattering from electron-phonon, and, as a result, our calculations show that $\rho_{\alpha\beta}\approx\rho^\uparrow_{\alpha\beta}$ with an order of magnitude difference in the conductivities $\sigma_{\alpha\beta}^\uparrow \gg \sigma^\downarrow_{\alpha\beta}$, while for Fe $\rho^\uparrow_{\alpha\beta}\sim\rho_{\alpha\beta}^\downarrow$. 
Since the spin up channel for Ni shows a larger lifetime broadening due to magnons~\cite{Muller2019}, the total lifetime $\tau_{n\mathbf{k}}^\uparrow$ is smaller than with electron-phonon scattering only, explaining the underestimated resistivity compared to experiment in Ni. In summary, comparing the calculated resistivity at low temperature with the experimental data of Fe and Ni, we find that the behavior of these two compounds is fundamentally different: magnons dominate the scattering in Ni while electron-phonon is the dominant scattering mechanism in Fe up to room temperature. 
We also consider the case of ferromagnetic Co HCP finding that the resistivity would be a case in between Fe and Ni; see Supplementary material S10~\cite{MCA2025}

\subsection{Role of magnetism on physical properties}
\begin{figure}[t]
  \centering
\includegraphics[width=0.95\linewidth]{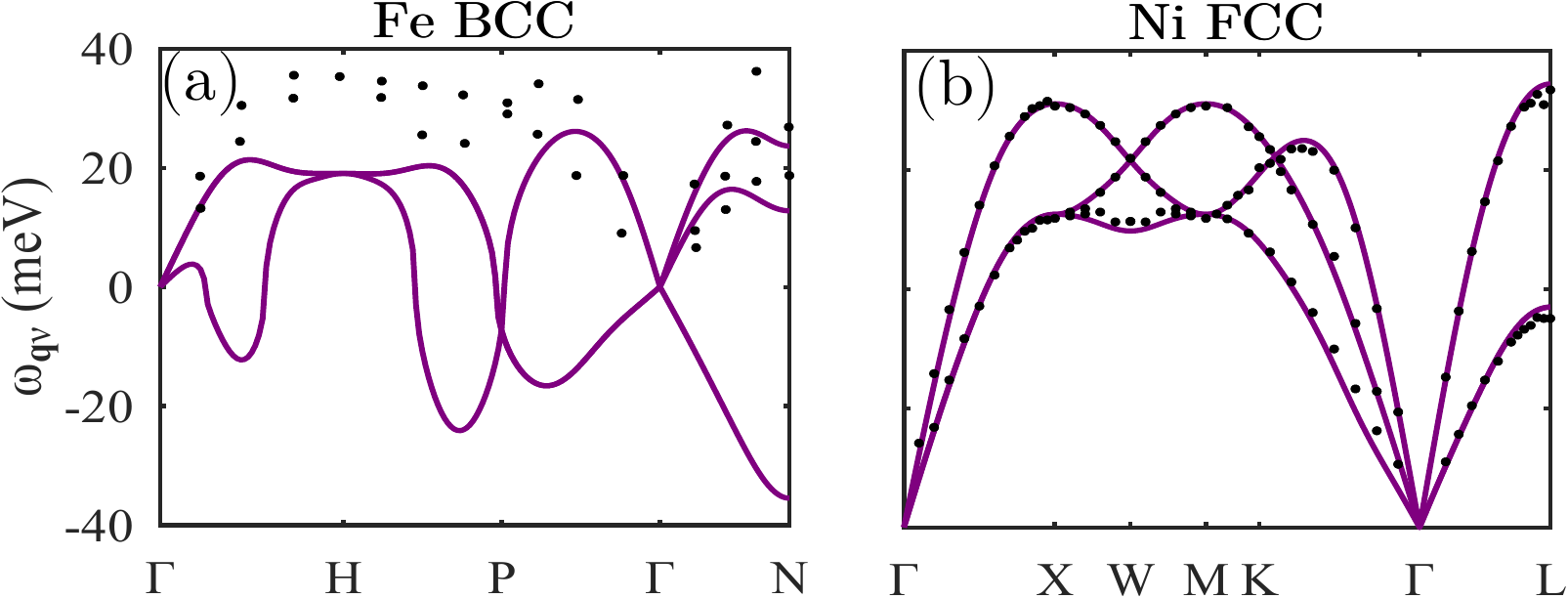}
    \caption{\label{Fe-Ni-ph-NM}
    Phonon dispersion for spin-unpolarized (non-magnetic) (a) Fe and (b) Ni obtained with PBE functional. 
    Black dots are the experimental data from Refs.~\cite{Birgeneau1964,Brockhouse1967}.
    }
\end{figure}

\begin{figure}[b]
    \centering
    \includegraphics[width=0.99\linewidth]{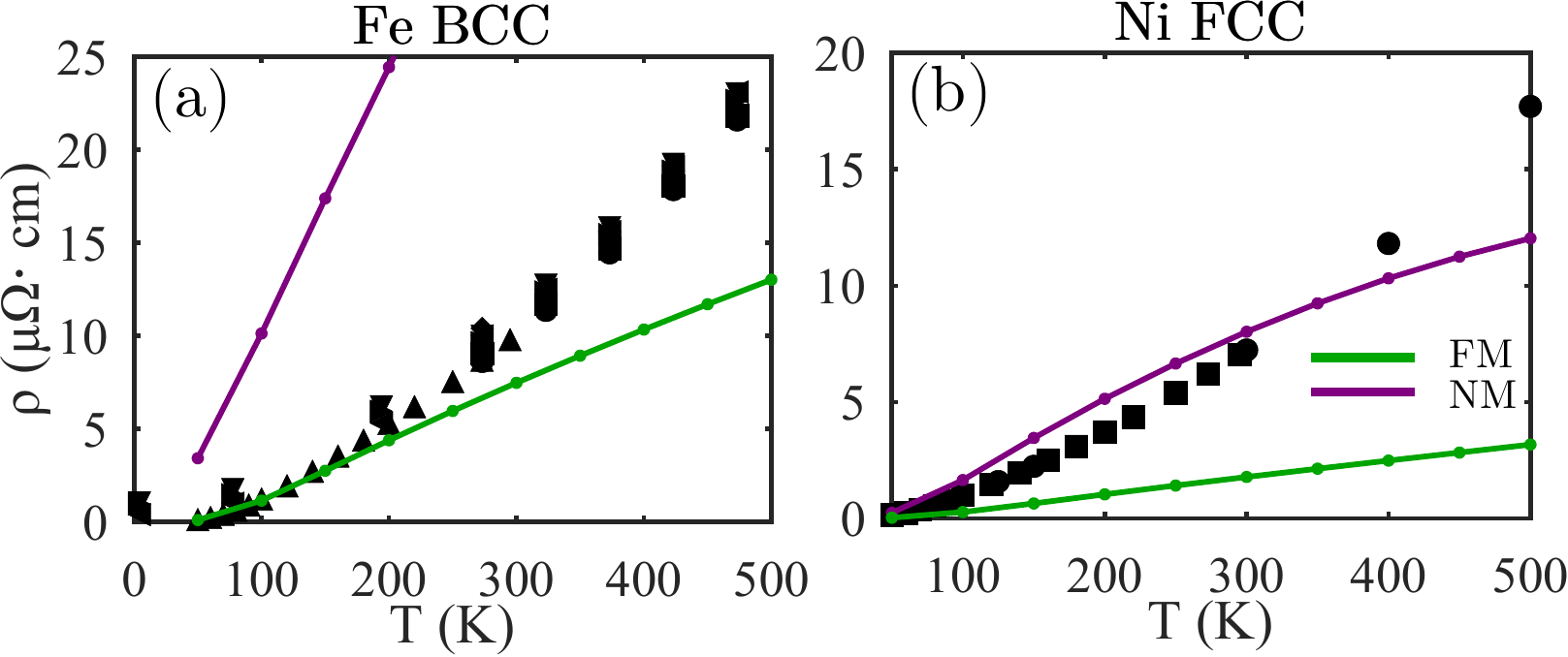}
    \caption{\label{NM-Fe-Ni-res}
    Electrical resistivity of iron and nickel as a function of temperature. 
    The solid green and purple lines are the computed PBE resistivities $\rho^{\rm BTE}$ using  ferromagnetic and non-magnetic EPC, respectively. 
    Black dots are experimental values obtained from Refs.~\cite{1959,Laubitz1976,Fulkerson1966}.    
    }
\end{figure}

One of the key aspects of including magnetism is the changes on both electronic and phonon band dispersion.
Therefore, we consider the effects of using the non-magnetic (NM) approximation in Fe and Ni by relaxing their BCC and FCC structures and calculating their respective phonon dispersions, see Fig.~\ref{Fe-Ni-ph-NM}. 
As we can see in Fig.~\ref{Fe-Ni-ph-NM}(a), Fe is not dynamically stable when we use the NM approximation. 
This is due to a fermionic pressure that increases the lattice parameters of Fe when magnetism is allowed~\cite{Janak1976Fe,Moruzzi1986,Marchant2019,Khemelevskyi2004}.
These instabilities found in the phonon dispersion for a BCC structure in the NM approximation are known to be related to distortions that lead to hexagonal phases~\cite{Grimvall2012}. 
This is also consistent with the fact that in the crystallographic phase diagram of Fe, we find a hexagonal phase at pressures a bit above 10~GPa~\cite{Cort1982,Taylor1991}. 
Interestingly, the phonon dispersion for Ni does not change much by using the NM approximation; see Fig.~\ref{Fe-Ni-ph-NM}(b).
In fact, one can notice that the FM phase of Ni produces only a small overall imbalance in the spin occupations since the experimental magnetic moment per unit cell is $M_{\mathrm{exp}} = 0.616~\mu_\mathrm{B}$~\cite{Danan1968} when compared with Fe that presents a magnetic moment of $M_{\mathrm{exp}} = (2.2 \pm 0.1)~\mu_\mathrm{B}$~\cite{vanAcker1988,Pasyuk1995,Bardos1969}.
Thus, one may wonder how different the resistivity in the NM phase is, when compared with that in the FM collinear case for Ni. 
In that regard, we repeat the calculations with the BTE for the resistivity of Ni with PBE (see LDA in the supplementary material S11~\cite{MCA2025}) but using the NM phase and we compare it in Fig.~\ref{NM-Fe-Ni-res}(b) with the FM case and with the experimental data.
We find that the NM approximation produces a slight overestimation of the resistivity for temperatures below 300~K and then it underestimates the resistivity. 

Although the limitation of the NM approximation has been highlighted previously~\cite{Varignon2019,Varignon2019b,Xiong2025,Xiong2025b,Malyi2023}, it is common to find studies that still use the NM approximation to model spin-disordered paramagnetic, the ferromagnetic or antiferromagnetic phases~\cite{Boeri2008,Meiser2024,Grissonanche2024}. In addition, some studies have also neglected the soft modes and still perform transport calculations~\cite{Zhang2017,Goudreault2025,Sohier2020}
For this reason, and in order to highlight how bad it would be to perform these calculations, we also compute the resistivity of NM Fe as a function of temperature and compare it with the FM case  in Fig.~\ref{NM-Fe-Ni-res}(a). 
The NM calculation with negative phonon frequencies yield an extremely overestimated resistivity when compared to the FM one and the experimental results.
Importantly, we show that even though the phonons are similar in Ni, neglecting the spin degree of freedom yields the wrong conclusion that the resistivity is dominated by electron-phonon interactions. 
We remark that, in spite of the seemingly better agreement with the experiment, the NM results actually overestimate the experimental data, which should never happen since we are not considering the full set of scattering channels.
Moreover, the temperature-dependent trend of the NM result is inconsistent with experimental observation. 
\section{Conclusions}\label{sec:conclusions}
We explore the role of electron-phonon interactions in typical magnetic metals. 
Namely, ferromagnetic Fe and Ni, where we show that neither the deformation potential nor the electron-phonon coupling constant present the same strength for the two spin channels, with the latter being dominated by Fermi surface nesting effects and the partial density of states of each spin channel.
Additionally, we show that the electrical resistivity of Fe and Ni exhibits important contributions from electron-phonon scattering at room temperature although the relative importance is material dependent. 
While in Fe the resistivity due to electron-phonon interactions is the dominant contribution at room temperature, for Ni it is less than one-third of the total. 
Additionally, we show that including the spin degree of freedom to compute resistivity is crucial. 
Neglecting it can lead to unstable phonons such as in the case of Fe, can overestimate resistivity, and can yield a completely erroneous interpretation for the source of the resistivity.
Furthermore, we explore the phonon-mediated superconducting mechanism in these two elemental compounds, a subject rarely addressed with proper numerical calculations.
We find suppressed superconductivity, showing that even in a non-magnetic configuration their critical temperatures would be extremely low.
These results validate and extend the capabilities of \textsc{EPW} to compute electron-phonon quantities with collinear magnetism. 
Beyond the methodological advances, our findings highlight how magnetism critically shapes resistive losses and superconductivity in metals that are central to energy applications. 
By enabling accurate modeling of transport in magnetic systems, this work provides a foundation for the design of low-energy-consumption spintronic devices, improved conductive pathways with reduced losses, and future magnetic materials for energy conversion and transmission. 
Future extensions to include spin–orbit coupling and non-collinear magnetism will further expand the reach of this framework, opening new directions for first-principles studies of complex magnetic materials.

\begin{acknowledgments}
S. P. is a Research Associate and M.M. is an Aspirant PhD candidate of the Fonds de la Recherche Scientifique - FNRS.
This work was supported by the Fonds de la Recherche Scientifique - FNRS under Grants number T.0183.23 (PDR). 
M.J.V. and G.A. acknowledge the Fonds de la Recherche Scientifique (FRS-FNRS Belgium) and Fonds Wetenschappelijk Onderzoek (FWO Belgium) for EOS project CONNECT (G.A. 40007563), and 
F\'ed\'eration Wallonie Bruxelles and ULiege for funding ARC project DREAMS (G.A. 21/25-11).
M.J.V. acknowledges funding by the Dutch Gravitation program
“Materials for the Quantum Age” (QuMat, reg number 024.005.006), financed by the Dutch Ministry of Education, Culture and Science (OCW).
Computational resources have been provided by the supercomputing facilities of the Université catholique de Louvain (CISM/UCL) and the Consortium des Équipements de Calcul Intensif en Fédération Wallonie Bruxelles (CÉCI) funded by the Fond de la Recherche Scientifique de Belgique (F.R.S.-FNRS) under convention 2.5020.11 and by the Walloon Region. 
The present research benefited from additional computational resources made available on Lucia, the Tier-1 supercomputer of the Walloon Region, infrastructure funded by the Walloon Region under the grant agreement n°1910247, and by EuroHPC (Extreme grant EHPC-EXT-2023E02-050) on Marenostrum5 at BSC, Spain.
\end{acknowledgments}

\bibliography{bibliography}
\end{document}


\renewcommand{\thefigure}{S\arabic{figure}}
\renewcommand{\thetable}{S\arabic{table}}
\renewcommand{\thesection}{S\arabic{section}}
\newcommand\SP[1]{{\color{blue}[SP:#1]}}

\title{Supplementary Material: Electron-phonon coupling in magnetic materials using the local spin density approximation}
\author{\'Alvaro Adri\'an Carrasco \'Alvarez}
\email{alvaro.carrasco@uclouvain.be}
\affiliation{European Theoretical Spectroscopy Facility, Institute of Condensed Matter and Nanosciences, Université catholique de Louvain, Chemin des Étoiles 8, B-1348 Louvain-la-Neuve, Belgium}
\author{Matteo Giantomassi}%
\affiliation{European Theoretical Spectroscopy Facility, Institute of Condensed Matter and Nanosciences, Université catholique de Louvain, Chemin des Étoiles 8, B-1348 Louvain-la-Neuve, Belgium}
\author{Jae-Mo Lihm}%
\affiliation{European Theoretical Spectroscopy Facility, Institute of Condensed Matter and Nanosciences, Université catholique de Louvain, Chemin des Étoiles 8, B-1348 Louvain-la-Neuve, Belgium}
\author{Guillaume E. Allemand}
\affiliation{European Theoretical Spectroscopy Facility, Nanomat/Q-Mat Universit\'e de Liège (B5), B-4000 , Liège Belgium}
\author{Maxime Mignolet}
\affiliation{European Theoretical Spectroscopy Facility, Nanomat Q-Mat University of Liège}
\author{Matthieu Verstraete}%
\affiliation{European Theoretical Spectroscopy Facility, Nanomat Q-Mat University of Liège}
\affiliation{ITP Dept of Physics, University of Utrecht, 3508 TA Utrecht, The Netherlands}
\author{Samuel Ponc\'e}%
\email{samuel.ponce@uclouvain.be}
\affiliation{European Theoretical Spectroscopy Facility, Institute of Condensed Matter and Nanosciences, Université catholique de Louvain, Chemin des Étoiles 8, B-1348 Louvain-la-Neuve, Belgium}
\affiliation{WEL Research Institute, avenue Pasteur 6, 1300 Wavre, Belgium.}
\date{\today}

\maketitle

\section{LOVA Resistivity formula}

One of the first formulas for computing the resistivity of metals due to electron phonon is the one derived by Ziman~\cite{Ziman1960} and later reformulated by Grimvall~\cite{Grimvall1981} that reads
\begin{equation}
    \rho^{\mathrm{LOVA}} = \frac{4\pi m_e}{n_e e^2k_\mathrm{B}T}\int d\omega\hbar\omega\alpha^2F_\mathrm{tr}(\omega)n(\omega,T)[n(\omega,T)+1]
    \label{rho}
\end{equation}
with $m_e$ the mass of the electrons, $n(\omega,T)$ the Bose-Einstein occupation factor, and $\alpha^2_\mathrm{tr}F(\omega)$ being the so called transport spectral function defined as:
\begin{multline}
    \alpha^2F_\mathrm{tr}(\omega) = \frac{1}{N(\varepsilon_\mathrm{F})} \sum_{\mathbf{k}\mathbf{q}}\sum_{nm\nu}w_\mathbf{k}w_\mathbf{q}|g_{mn\nu}(\mathbf{k},\mathbf{q})|^2\delta(\omega-\omega_{\mathbf{q}\nu}) \\
    \times \Big[1-\frac{\mathbf{v}_{n\mathbf{k}}\cdot\mathbf{v}_{m\mathbf{k+q}}}{|\mathbf{v}_{n\mathbf{k}}|}\Big]\delta(\varepsilon_{n\mathbf{k}}-\varepsilon_{\mathrm{F}})\delta(\varepsilon_{m\mathbf{k+q}}-\varepsilon_{\mathrm{F}}),
\end{multline}
where $N(\varepsilon_\mathrm{F})=\sum_n\int\frac{\mathrm{d^3\mathbf{k}}}{\Omega^\mathrm{BZ}}\delta(\varepsilon_{n\mathbf{k}}-\varepsilon_\mathrm{F})$ is the total density of states at the fermi level.  
Interestingly, this formula accounts for spin orbit coupling (SOC) in the Hamiltonian with $\sum_{\mathbf{k}} w_\mathbf{k} = \sum_{\mathbf{q}} w_\mathbf{q} = 1$. 
If we neglect SOC, the Hamiltonian is diagonal and the spin index $\sigma$ is a good quantum number and we obtain: 
\begin{multline}
    \alpha^2F_\mathrm{tr}(\omega) = \sum_{\sigma} \frac{1}{N(\varepsilon_\mathrm{F})}\sum_{\mathbf{k}\mathbf{q}}\sum_{nm\nu}w_\mathbf{k}w_\mathbf{q}|g^\sigma_{{m}{n}\nu}(\mathbf{k},\mathbf{q})|^2\\\times\left(1-\frac{\mathbf{v}^\sigma_{{n}\mathbf{k}}\cdot\mathbf{v}^\sigma_{{m}\mathbf{k+q}}}{|\mathbf{v}^\sigma_{{n}\mathbf{k}}|}\right)\delta(\varepsilon^\sigma_{{n}\mathbf{k}}-\varepsilon_{\mathrm{F}})\delta(\varepsilon^\sigma_{{m}\mathbf{k+q}}-\varepsilon_{\mathrm{F}})\delta(\omega-\omega_{\mathbf{q}\nu}), 
\end{multline}
where the $n$ and $m$ band indices are restricted to the specific spin subspace and $\sum_{n} \rightarrow \sum_{n\sigma}$.
We can then define a spin resolved transport spectral function $\alpha^2F_\mathrm{tr}^\sigma$ such that $\alpha^2F_\mathrm{tr}(\omega)$ is additive in the spin indices $\sigma$
\begin{equation}\label{eq:trans}
    \alpha^2F_\mathrm{tr}(\omega) = \sum_\sigma\alpha^2F_\mathrm{tr}^\sigma(\omega) = \alpha^2F^\uparrow_\mathrm{tr}(\omega) + \alpha^2F^\downarrow_\mathrm{tr}(\omega).
\end{equation}
By introducing Eq.~\eqref{eq:trans} into Eq.~\eqref{rho}, we have that $\rho^{\mathrm{LOVA}}$ is also additive in the spin indices $\sigma$
\begin{equation}
    \rho^{\mathrm{LOVA}} = \tilde{\rho}^{\uparrow} + \tilde{\rho}^\downarrow,
\end{equation}
which is similar to a series resistor model for the 2 spin channels.
Interestingly, we note that this additive property is different from $\rho^\mathrm{BTE}$ defined in Eqs.~(13) and (14) of the manuscript:
\begin{equation}
    (\rho^\mathrm{BTE} )^{-1}= (\rho^\uparrow)^{-1} + (\rho^\downarrow)^{-1},
\end{equation}
which is a parallel resistor model for the 2 spin channels. 
We note that $\rho^{\mathrm{LOVA}} \approx \rho^\mathrm{BTE}$ and describe the same observable while $\rho^\uparrow/\rho^\downarrow$ and $\tilde{\rho}^\uparrow/\tilde{\rho}^\downarrow$ do not.

The reason for this is that the spin resolved resistivity obtained with the BTE can be interpreted as the resistivity that each spin channel would experience and is derived in the manuscript.
Instead, the spin resolved resistivity obtained from $\alpha^2F_\mathrm{tr}^\sigma$ leads to $\tilde{\rho}^\uparrow$ and $\tilde{\rho}^\downarrow$
which are relative contributions to the total resistivity of each spin channel. 

This difference can be traced back to the derivation of Ziman~\cite{Ziman1960} where the resistivity is connected to the change of entropy $\dot{S}$  due to an electric current $\mathbf{J}$ at unit electric field $\mathbf{E}$ and given as $\dot{S} / (\mathbf{J}^2 T)$.
Under this interpretation, it is natural to find an additive definition of resistivity with respect to the spin channels, since the entropy of the system is an extensive quantity and therefore additive $(\dot{S}^{\uparrow}+\dot{S}^\downarrow)/((\mathbf{J^{\uparrow}+\mathbf{J}^\downarrow})^2T)$.

\section{Verification of the electrical resistivity in magnetic compounds}
\begin{figure}[t]
    \centering
    \includegraphics[width=0.99\linewidth]{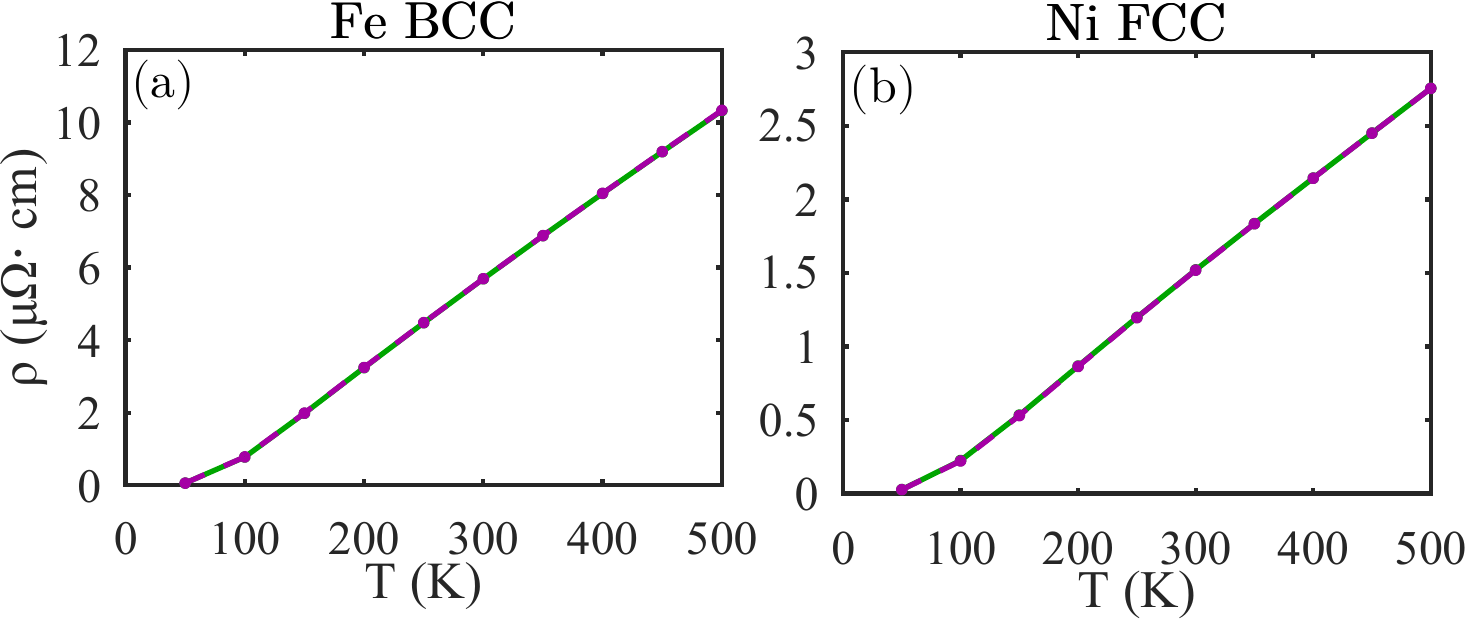}
    \caption{Electrical resistivity as a function of temperature for ferromagnetic iron (a) and nickel (b) computed with LDA. 
    Green solid lines corresponds to \textsc{EPW} calculations while purple dashed lines corresponds to \textsc{ElectronPhonon.jl}.}
    \label{JL-res}
\end{figure}

We verify the results of the implementation of the Boltzmann transport equation (BTE) in \textsc{EPW} with other codes. 
Since no public codes provide the BTE for magnetic materials, we extend the capability of the \textsc{ElectronPhonon.jl} package~\cite{Lihm2024,EPjl} for the purpose of this work. 
We compare the resistivity of Fe and Ni on a 112$\times$112$\times$112 \textbf{k}- and \textbf{q}-point grids in Fig.~\ref{JL-res} between \textsc{EPW} and \textsc{ElectronPhonon.jl}, finding perfect agreement (largest difference is 2\%).
We note that the \textsc{ElectronPhonon.jl} package relies on the real-space electron-phonon matrix elements provided by \textsc{EPW}, thereby providing only a verification of the interpolation part of the calculations. 
We also report that the \textsc{ElectronPhonon.jl} code is about three times faster due to an inversion of the electron and phonon momentum grids loops.
However, \textsc{ElectronPhonon.jl} currently uses Julia's native multithreading to parallelize instead of the MPI approach employed in \textsc{EPW}
and is therefore limited to a single compute node.

\section{Recovering the non-magnetic approximation in $\rm Pb$}
\begin{figure}[b]
    \centering
    \includegraphics[width=0.99\linewidth]{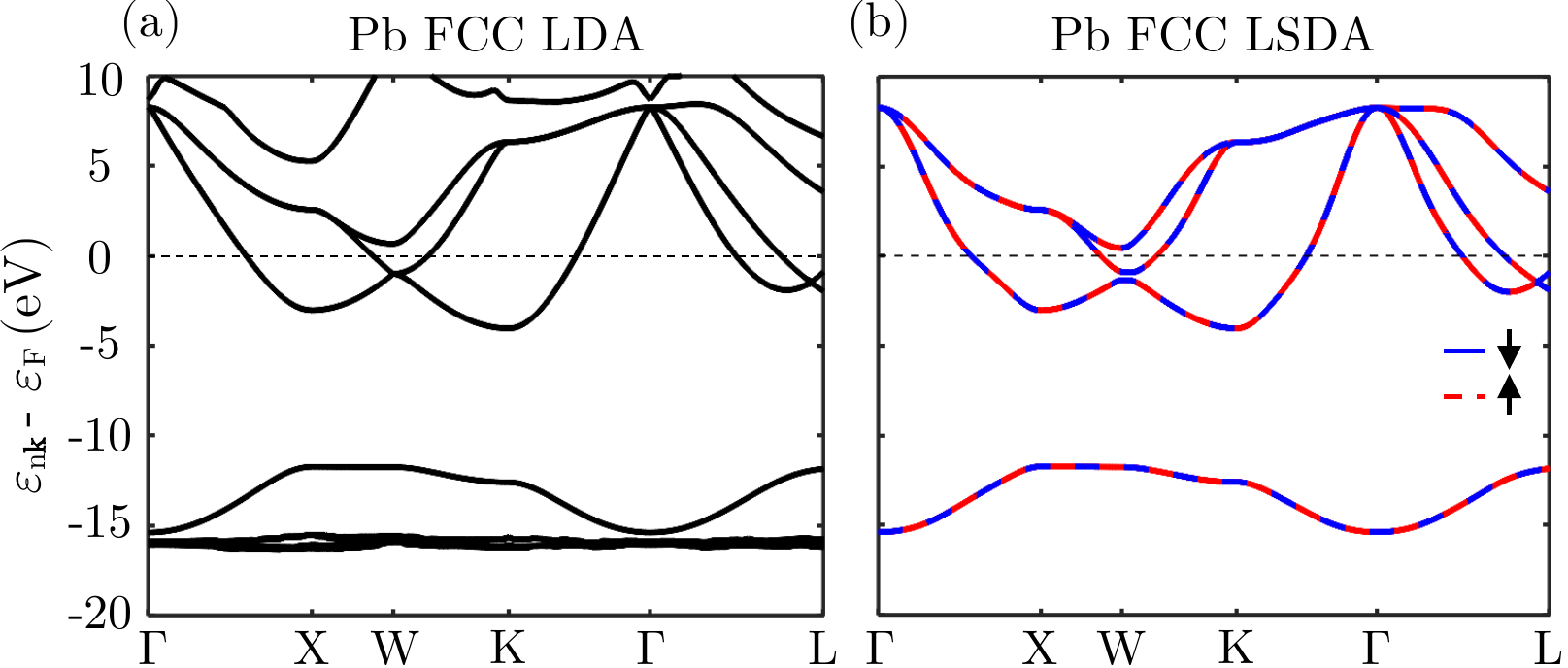}
    \caption{Electronic band structure of Pb with (a) LDA computed with \textsc{QE} and (b) LDSA interpolated with \textsc{EPW}.}
    \label{Pb-LSDA-bands}
\end{figure}

We compute the electronic dispersion and the deformation potential of Pb using LDA and LSDA. 
Since Pb is a Pauli paramagnet, the spin density for the spin up and spin down channels will be equivalent at every point in space:
\begin{equation}
    \rho^\uparrow(\mathbf{r}) = \rho^\downarrow(\mathbf{r}), \; \forall\,\mathbf{r} \in \mathbb{R}^3.
\end{equation}
This implies that all interpolated quantities for the two spin channels must be identical up to numerical precision when the same set of parameters such as $\mathbf{k}$ grids and cutoff energy are used for the calculation with LDA and LSDA. 
We confirm this in Fig.~\ref{Pb-LSDA-bands} for the electronic band structure by comparing non-magnetic LDA with collinear magnetic LSDA.
We find that the electronic band structures produce the same results for both spin channels (machine precision) and 
are close to the non-magnetic LDA calculation (largest difference below 0.02\%).
\begin{figure}[t]
    \centering
    \includegraphics[width=0.65\linewidth]{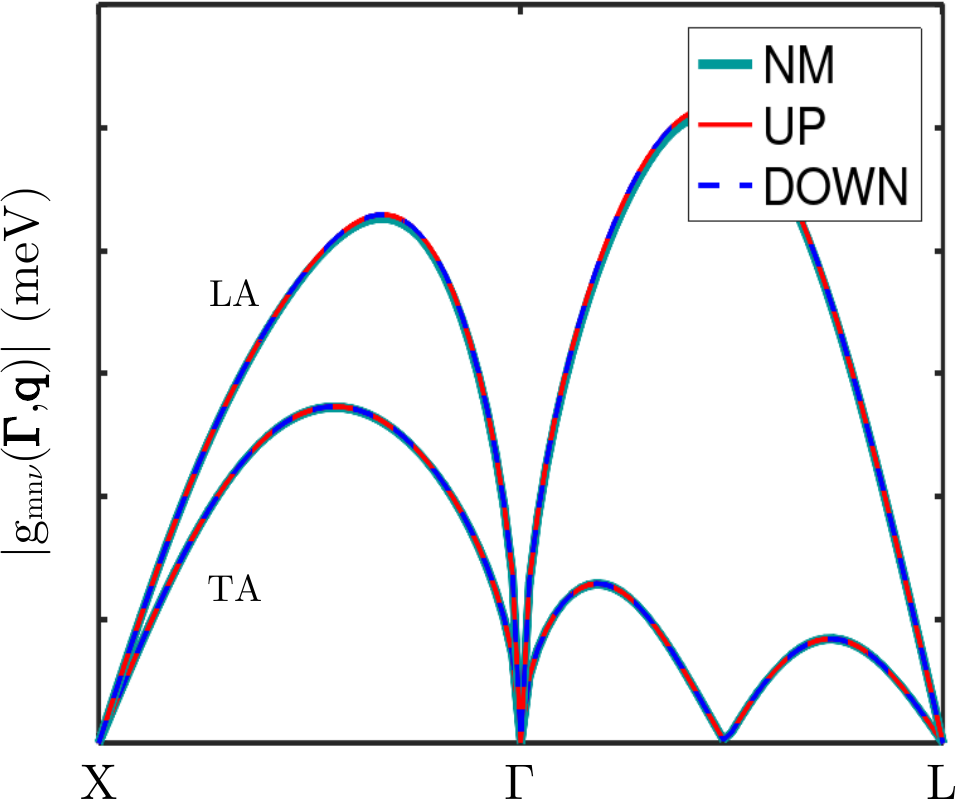}
    \caption{\label{Pb-LSDA-g}
    Absolute value of the electron phonon matrix elements for bands $m = n = 8$ with LDA (non magnetic) and LSDA (up and down channels), compared to the non-magnetic (NM) solution.}
\end{figure}
These results show that the interpolation for both spin channels works as expected and the non-magnetic solution is recovered.
In addition, we compare the interpolated electron-phonon coupling $g_{mn\nu}(\mathbf{k},\mathbf{q})$ in Fig.~\ref{Pb-LSDA-g}, obtaining excellent agreement (largest difference below 0.03\%).

\begin{figure}[t]
    \centering
    \includegraphics[width=0.65\linewidth]{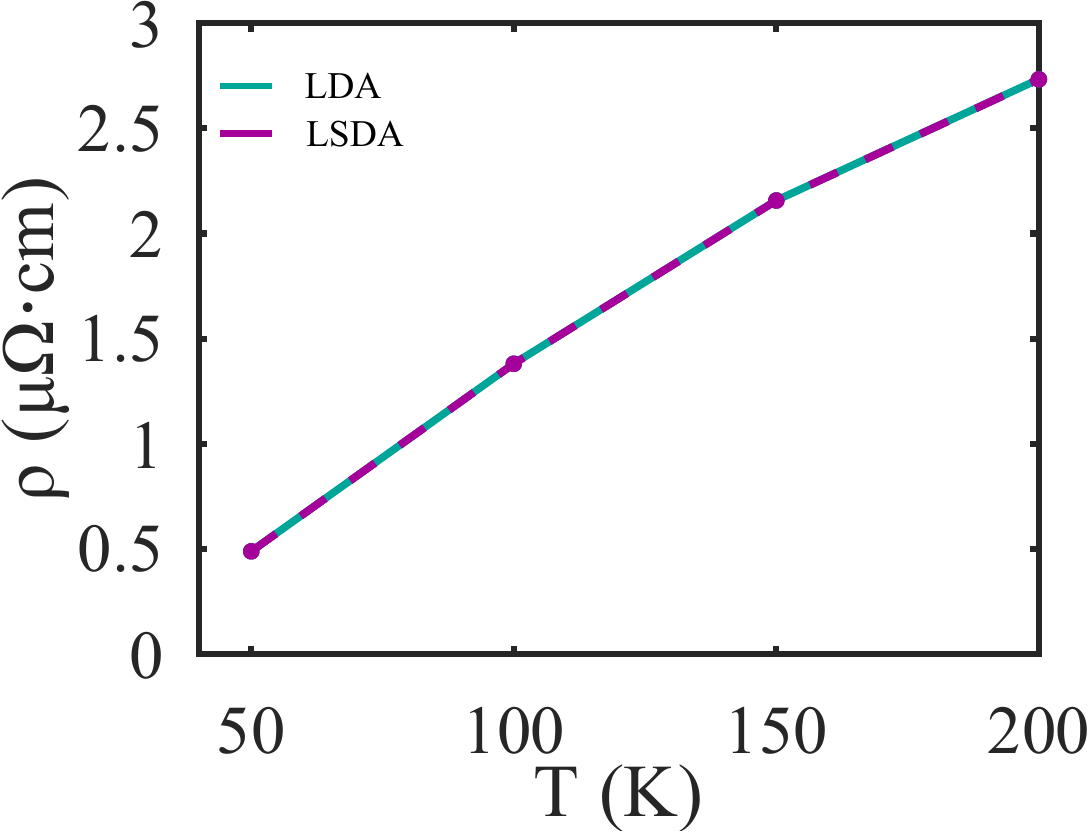}
    \caption{\label{Pb-LSDA-LDA-res}
    Computed resistivity of FCC Pb using LDA (purple) and LSDA (green).}
\end{figure}

Finally, we compute in Fig.~\ref{Pb-LSDA-LDA-res} the resistivity of Pb with the BTE with LDA and LSDA, also obtaining a good agreement  (largest difference below 0.04\%).

\section{Verification of \textsc{QE} and \textsc{Abinit} for ferromagnetic $\rm Fe$ and $\rm Ni$}
We report the lattice parameters and the total magnetization computed with \textsc{Quantum ESPRESSO} and \textsc{Abinit} in Table~\ref{Table}.

{\renewcommand{\arraystretch}{0.9}
\begin{table*}[t]
    \centering
    \begin{tabular}{l l l l l l l l l }
    \hline
      Code & Compound & $a$ (Bohr) & Magnetic order & $M^{\mathrm{Tot}}$ ($\mu_{\mathrm{B}}$) & Functional & E$^{\mathrm{cut}}$ (Ry)   & $\bf k$-mesh & Pseudopotential  \\ \hline
    QE   &  Fe            &  5.21   & NM                 & 0       & PBE & 100 & 24$^3$ & PseudoDojo v0.5 stand.\\
         &                &  5.36   & FM - collinear     & 2.26   & PBE & 100 & 24$^3$ & PseudoDojo v0.5 stand.\\ 
         &                &  5.09   & NM                 & 0       & LDA & 100 & 24$^3$ & PseudoDojo v0.5 stand. \\
         &                &  5.20   & FM - collinear     & 2.05   & LDA & 100 & 24$^3$ & PseudoDojo v0.5 stand. \\ \hline
\textsc{Abinit} &         &  5.21  & NM                  & 0      & PBE & 100 & 24$^3$ & PseudoDojo v0.5 stand.\\
                &          &  5.36  & FM - collinear      & 2.25   & PBE & 100 & 24$^3$ & PseudoDojo v0.5 stand.\\ 
                &          &  5.09  & NM                  & 0      & LDA & 100 & 24$^3$ & PseudoDojo v0.5 stand. \\
                &          &  5.20  & FM - collinear      & 2.04   & LDA & 100 & 24$^3$ & PseudoDojo v0.5 stand. \\\hline         
Experiment &              &  5.42~\cite{Fe-Ni-latpar}   & FM     & 2.22~\cite{Fe-mag1,Fe-mag2,Fe-mag3,Fe-Ni-magmom}  & -   & -      & - & -  \\ \hline
    QE   &  Ni            &  6.64   & NM                  & 0       & PBE & 100 & 24$^3$ & PseudoDojo v0.5 stand. \\
         &                &  6.65   & FM - collinear      & 0.65   & PBE & 100 & 24$^3$ & PseudoDojo v0.5 stand. \\ 
         &                &  6.46   & NM                   & 0       & LDA & 100 & 24$^3$ & PseudoDojo v0.5 stand. \\
         &                &  6.47   & FM - collinear      & 0.57   & LDA & 100 & 24$^3$  & PseudoDojo v0.5 stand. \\ \hline
\textsc{Abinit} &        &  6.64  & NM                  & 0      & PBE & 100 & 24$^3$ & PseudoDojo v0.5 stand. \\
                &          &  6.65  & FM - collinear       & 0.67   & PBE & 100 & 24$^3$ & PseudoDojo v0.5 stand. \\ 
                &          &  6.46  & NM                  & 0      & LDA & 100 & 24$^3$  & PseudoDojo v0.5 stand. \\
                &          &  6.47  & FM - collinear       & 0.57   & LDA & 100 & 24$^3$ & PseudoDojo v0.5 stand. \\          
Experiment &              &  6.66~\cite{Fe-Ni-latpar}   & FM       & 0.62~\cite{Fe-Ni-magmom}  & -   & -      & - & - \\ \hline  
    \end{tabular}
    \caption{\label{Table} Comparison between \textsc{Quantum ESPRESSO} (QE) and \textsc{Abinit} for the lattice parameter $a$ and the total magnetization $M^{\mathrm{Tot}}$ as well as computational parameters including planewave energy cutoff E$^{\mathrm{cut}}$, $\textbf{k}$-point grid and pseudopotential from \textsc{PseudoDojo}~\cite{vanSetten2018}.}
\end{table*}
}

\begin{figure}[tb]
    \centering
    \includegraphics[width=0.99\linewidth]{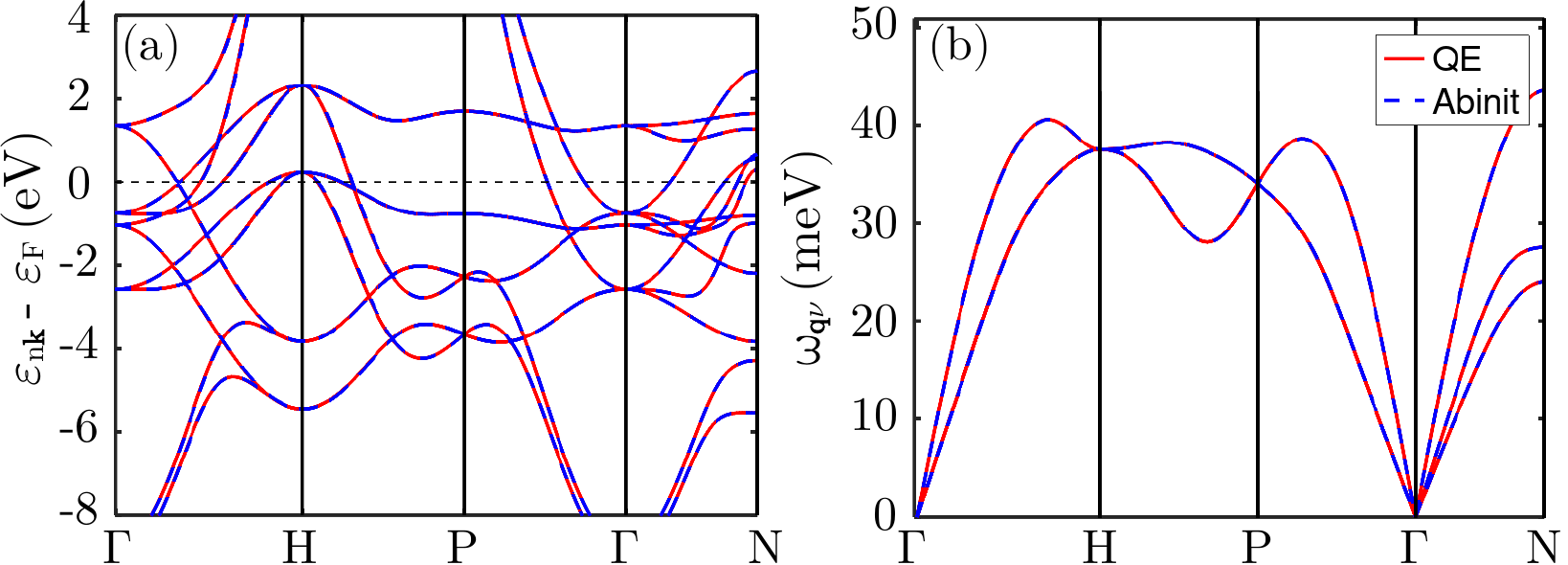}
    \caption{Electronic band structure (a) and phonon dispersion (b) of ferromagnetic Fe with \textsc{Quantum ESPRESSO (QE)}  and \textsc{Abinit}.}
    \label{QE-Abinit-bands}
\end{figure}

We find excellent agreement between the two codes with a difference between the total energy below 10$^{-5}$ Ry, similar to previous studies without magnetism~\cite{ponce2014verification,Ponce2025}.
Such agreement is obtained by carefully using the same pseudopotential, FFT grid, and cutoff radius for the non-local part of the potential. 
We also compare the electronic bandstructure and phonon dispersion in Fig.~\ref{QE-Abinit-bands} and find differences as small as previously reported with non- magnetic compounds~\cite{ponce2014verification}.

\section{Verification of the electron-phonon matrix element between \textsc{Quantum ESPRESSO} and \textsc{Abinit}}

\begin{figure}[t]
    \centering
    \includegraphics[width=0.99\linewidth]{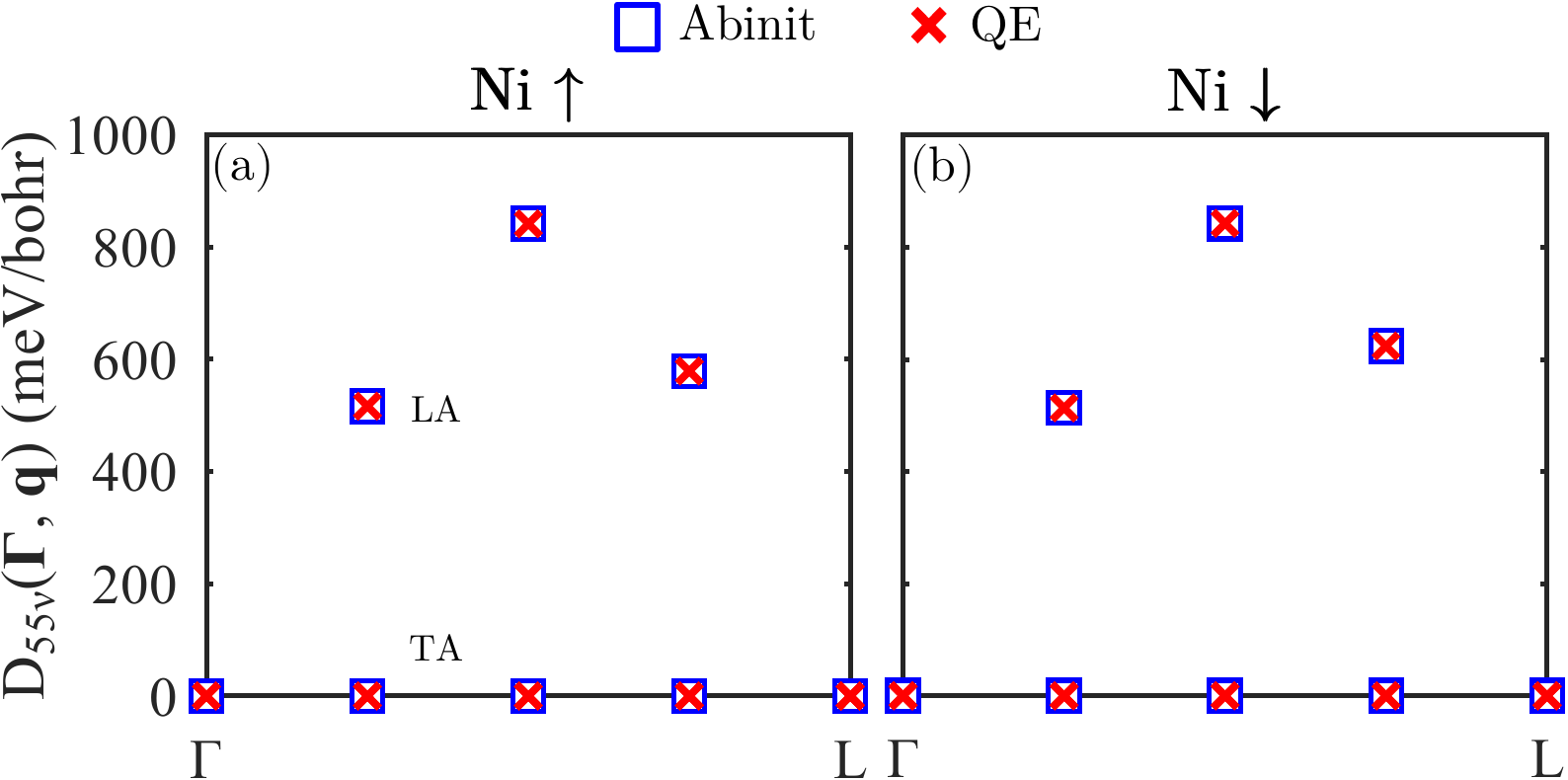}
    \caption{Direct evaluation of the deformation potential of ferromagnetic Ni for (a) spin up and (b) spin down at $\mathbf{k} = \mathbf{\Gamma}$ for $m = n = 5$, along a $\mathbf{q}$ momentum line for all phonon modes $\nu$. 
    Red dots are computed with \textsc{Quantum ESPRESSO} (QE) while blue squares are computed with \textsc{Abinit}. LDA functional was employed.}
    \label{D-Fe-Ni}
\end{figure}

We compare in Fig.~\ref{D-Fe-Ni} the absolute value of the directly calculated electron-phonon matrix element between \textsc{Quantum ESPRESSO} and \textsc{Abinit} of ferromagnetic Ni at $\mathbf{k} = \Gamma$ for $m = n = 5$, which is one of the bands crossing the Fermi level.
The results show good agreement, with the largest difference being 0.01~meV/bohr.
\begin{figure}[t]
    \centering
    \includegraphics[width=0.99\linewidth]{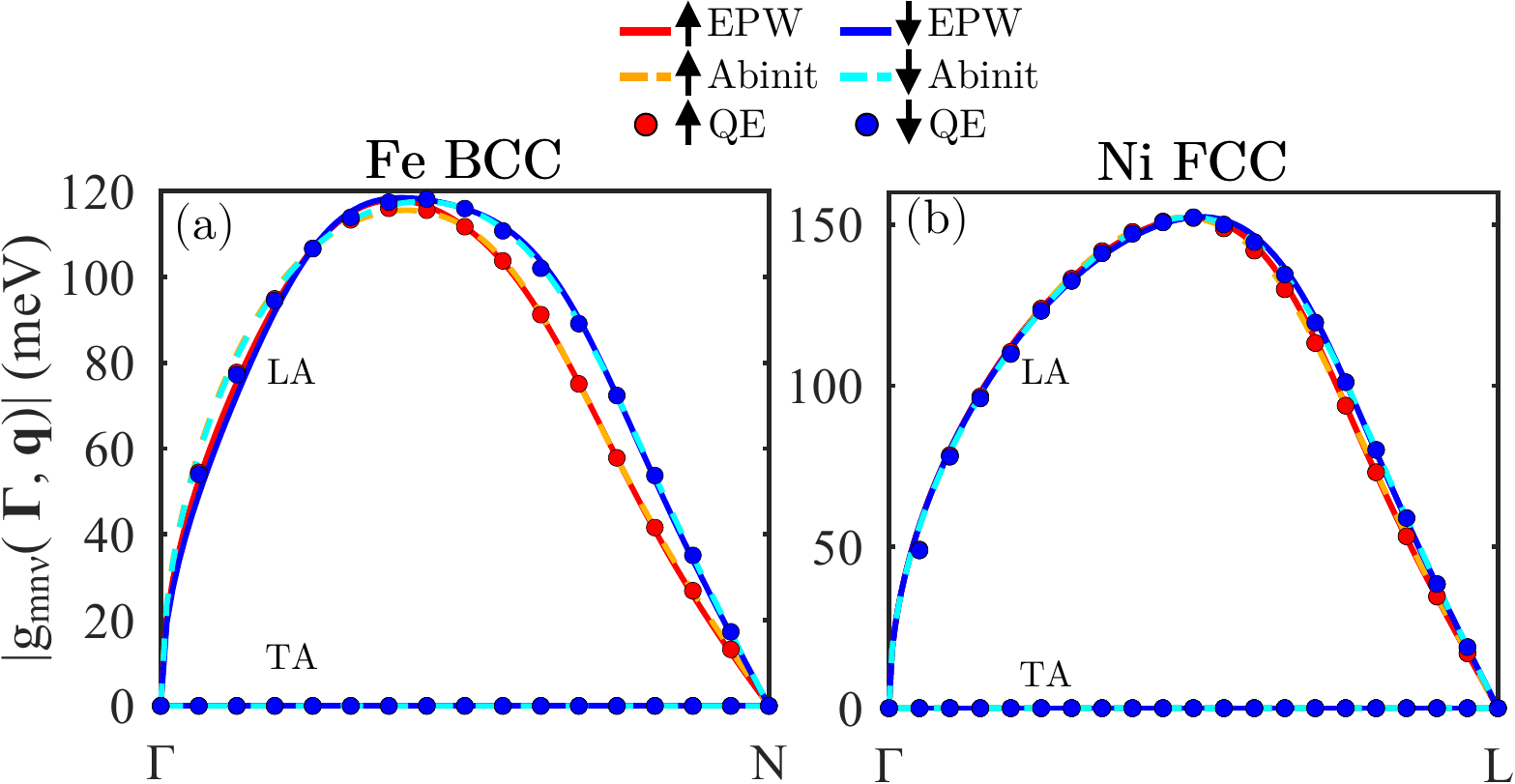}
    \caption{\label{g-Fe-Ni-qe-abinit}
    Electron phonon matrix element for (a) Ferromagnetic Fe, and (b) Ferromagnetic Ni at $\mathbf{k} = \mathbf{\Gamma}$ for $m = n = 5$, along a $\mathbf{q}$ momentum line for all phonon modes $\nu$. 
    Red dots are computed directly with \textsc{Quantum ESPRESSO} (QE) while the solid/dashed lines are interpolated with \textsc{EPW}/\textsc{Abinit}. LDA functional was employed.}
\end{figure}

Additionally, we also compare the interpolated electron phonon matrix elements between \textsc{Abinit} and \textsc{EPW} using the same 8$\times$8$\times$8 $\mathbf{q}$ grids for the phonon part and for the Wannier interpolation of the electronic structure in \textsc{EPW} we use an 8$\times$8$\times$8 $\mathbf{k}$ grid. 
The results are shown in Fig.~\ref{g-Fe-Ni-qe-abinit}(a) and Fig.~\ref{g-Fe-Ni-qe-abinit}(b) for Fe and Ni where we can find a similar behavior of the interpolation methods of \textsc{EPW} and \textsc{Abinit}, although not completely equivalent close to the $\boldsymbol{\Gamma}$ point.

Interestingly, we note that the spin up and down electron-phonon matrix elements are visibly different for the case of iron but close in the case of nickel.

\section{Real space decay of interpolated quantities}

\begin{figure}[b]
    \centering
    \includegraphics[width=0.99\linewidth]{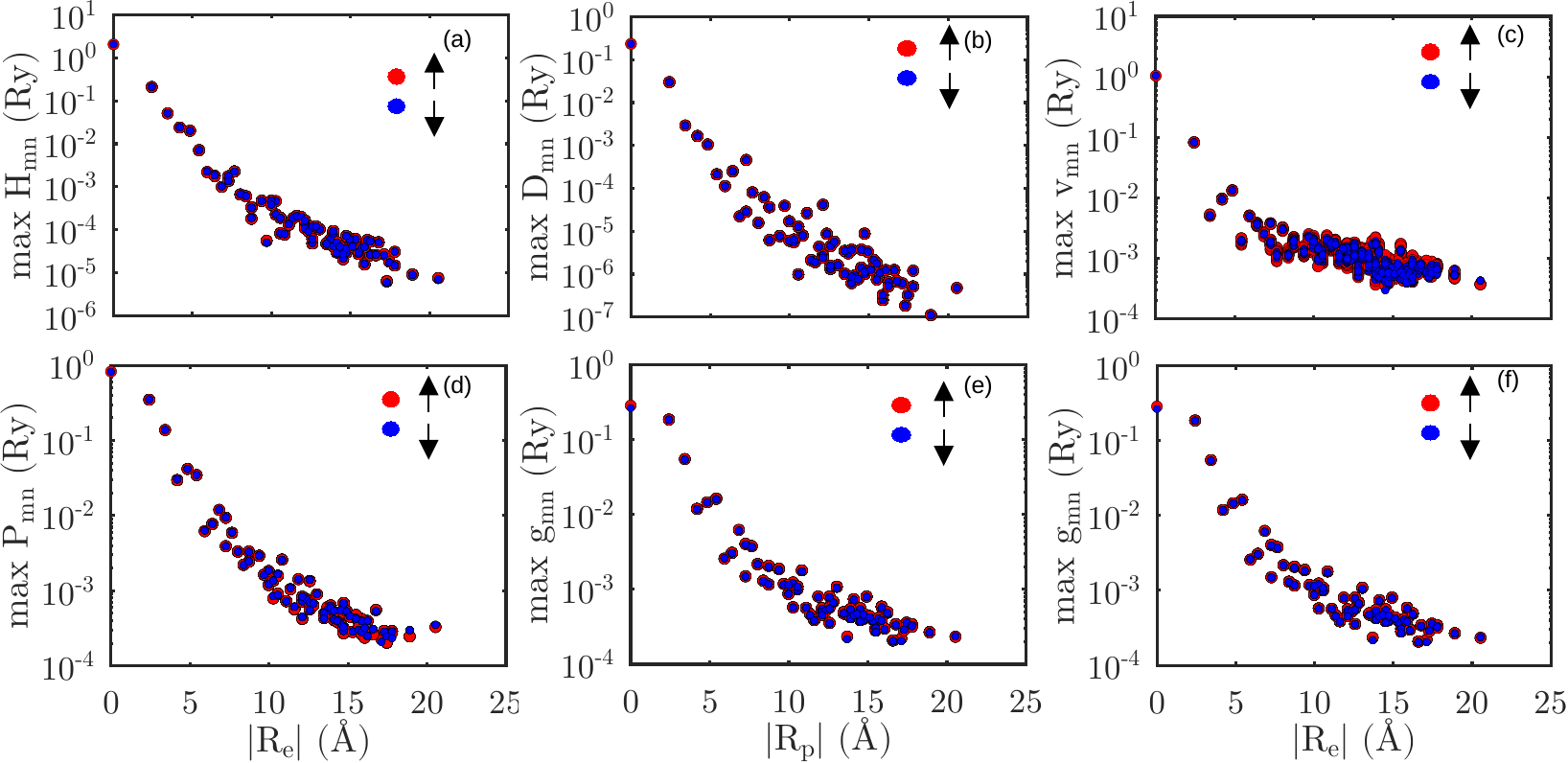}
    \caption{\label{Decay}
    Real space decay of the Hamiltonian (a), dynamical matrix (b), velocity matrix element (c), dipole matrix element (d), and electron phonon matrix element in the limiting cases $\mathbf{R}_e$ = 0 (e) and $\mathbf{R}_p = 0$ (f). }
\end{figure}

For an accurate interpolation, the Hamiltonian, dynamical matrices and electron-phonon matrix elements should be localized in real-space. 
We show in Fig.~\ref{Decay} the decay of these quantities in real space and find that both spin channels decay in a similar way.

We present the Hamiltonian in Fig.~\ref{Decay}(a), the dynamical matrix in (b), the velocity matrix elements in (c), the dipole matrix elements in (d) and the electron phonon matrix elements in real space $g(\mathbf{R}_e,\mathbf{R}_{p})$ for the limiting cases of $\mathbf{R}_e = 0$ and $\mathbf{R}_{p}$ in (e) and (f) respectively.

In all cases, we find a decay of at least three orders of magnitude, confirming the quality of the interpolation. 

\section{Electron phonon coupling strength and superconductivity}

Phonon-mediated superconductivity has been studied with different formalisms that often do not consider magnetism. 
In particular, the critical temperature of a superconductor can be computed using the Allen-Dynes formula~\cite{Allen1975}:
\begin{equation}
  T_{\rm c}^{\rm AD} = \frac{\hbar\omega_{\rm log}}{1.2 k_{\rm B}}\mathrm{exp}{\left(\frac{-1.04(1+\lambda)}{\lambda - \mu^*(1+0.62\lambda)}\right)},
\end{equation}
where $\lambda$ is the electron phonon coupling strength, $k_{\rm B}$ is the Boltzmann constant,  $\mu^*$ is the semi-empirical Coulomb potential, and $\omega_{\rm log}$ is the logarithmic average of the phonon frequency (geometric average) defined as
\begin{equation}
    \omega_{\rm log} \equiv \mathrm{exp}\left(\frac{1}{\lambda}\int\frac{d\omega}{\omega}\alpha^2F(\omega)\log{\omega}\right),
\end{equation}
with $\alpha^2F(\omega)$ being the Eliashberg spectral function. 
Although this formula is generally used for the $s$-wave pairing mechanism with time reversal symmetry, a BCS type formula can be obtained for the other gap channels $p$, $d$, or $f$ with spin singlet and triplet~\cite{Anderson1961,Houzet2012} which requires form factors and different semi-empirical numerical parameters.

However, some authors have used the standard Allen-Dynes formula in the context of phonon mediated superconductivity in magnetic systems~\cite{Zhang2025}, where $\lambda$ is identified with the average mass enhancement parameter. 
Interestingly, the use of the standard formula is valid when considering an $s$-wave singlet in a commensurate collinear anti-ferromagnet (AFM) with inversion symmetry and spin degeneracy on the bands such that $\varepsilon_{n\mathbf{k}}^\uparrow = \varepsilon_{n\mathbf{k}}^\downarrow$ and  $g_{mn\nu}^\uparrow(\mathbf{k},\mathbf{q}) = g^\downarrow_{mn\nu}(\mathbf{k},\mathbf{q})$ as done in Ref.~\cite{Zhang2025}.

Another valid use of the standard Allen-Dynes formula is for $s$-wave triplet ($S_z = \pm1$) with interband pairing involving transitions $n, \mathbf{k}\rightarrow m, \mathbf{k+q}$ and $m,\mathbf{-k}\rightarrow n, \mathbf{-k-q}$. However, we must note that the $\lambda$ for the pairing and the mass enhancement are not necessarily equivalent. While the mass enhancement is proportional to $\lambda_{\mathrm{mass}}\propto|g^\sigma_{mn\nu}(\mathbf{k},\mathbf{q})|^2$, the pairing $\lambda$ would be proportional to $\lambda_{\mathrm{pair}}\propto g^\sigma_{mn\nu}(\mathbf{k},\mathbf{q})g^\sigma_{mn\nu}(\mathbf{-k},\mathbf{-q})$. In the case of the $s$-wave pairing, we can consider that only the real part of the matrix elements are involved, $\lambda_\mathrm{pair} \propto \mathrm{Re}[g^\sigma_{mn\nu}(\mathbf{k},\mathbf{q})g^\sigma_{nm\nu}(\mathbf{-k},\mathbf{-q})]$. Under this constrains, it is interesting to consider the following inequality
\begin{multline}
    2\mathrm{Re}[g^\sigma_{mn\nu}(\mathbf{k},\mathbf{q})g^\sigma_{nm\nu}(-\mathbf{k},\mathbf{-q})]\\\leq 
    |g^\sigma_{mn\nu}(\mathbf{k},\mathbf{q})|^2 + |g^\sigma_{nm\nu}(\mathbf{-k},\mathbf{-q})|^2.
\end{multline}
If we now divide by $\mathbf{\omega_{\mathbf{q}\nu}}$ and $N(\varepsilon_\mathrm{F})$, and sum over all $\mathbf{k}$ and $\mathbf{q}$ such that they belong to the Fermi surface pairs ($\mathrm{F.S.}_\mathrm{pair}$) $\mathbf{k}$ and $\mathbf{-k}$ as well as $\mathbf{k+q}$ and $\mathbf{-k-q}$, and sum over bands $m$, $n$, and $\nu$. We obtain:
\begin{multline}
    \sum_{mn\nu}\sum_{\mathbf{k},\mathbf{q}\in \mathrm{F.S.}_\mathrm{pair}}\frac{2\mathrm{Re}[g^\sigma_{mn\nu}(\mathbf{k},\mathbf{q})g^\sigma_{nm\nu}(\mathbf{-k},\mathbf{-q})]}{\omega_{\mathbf{q}\nu}N(\varepsilon_\mathrm{F})}\\\leq
    \sum_{mn\nu}\sum_{\mathbf{k},\mathbf{q}\in \mathrm{F.S.}_\mathrm{pair}}\frac{|g^\sigma_{mn\nu}(\mathbf{k},\mathbf{q})|^2 + |g^\sigma_{nm\nu}(\mathbf{-k},\mathbf{-q})|^2}{\omega_{\mathbf{q}\nu}N(\varepsilon_\mathrm{F})}\\\leq
    \sum_{mn\nu}\Bigg(\sum_{\mathbf{k},\mathbf{q}\in \mathrm{F.S.}}\frac{|g^\sigma_{mn\nu}(\mathbf{k},\mathbf{q})|^2}{\omega_{\mathbf{q}\nu}N(\varepsilon_\mathrm{F})} + \\\sum_{\mathbf{-k},\mathbf{-q}\in \mathrm{F.S.}}\frac{|g^\sigma_{nm\nu}(\mathbf{-k},\mathbf{-q})|^2}{\omega_{\mathbf{q}\nu}N(\varepsilon_\mathrm{F})}\Bigg) \iff
    2\lambda^\sigma_\mathrm{pair}\leq 2\lambda^\sigma_\mathrm{mass},
\end{multline}
where in the last term we have used the fact that $\sum_{\mathbf{k}\in\mathrm{F.S.}_\mathrm{pair}}\leq \sum_{\mathbf{k}\in\mathrm{F.S.}}$where $\mathrm{F.S.}$ indicates the whole Fermi surface. This result shows that $\lambda^\sigma_{\mathrm{mass}}\geq \lambda^\sigma_\mathrm{pair}$ for this type of interband pairing. We note that the equality $\lambda^\sigma_\mathrm{mass}=\lambda^\sigma_\mathrm{pair}$ would only hold for systems with time reversal symmetry. Thus, we can conclude that the mass enhancement can give us an upper bound for this type of $s$-wave triplet pairing where time reversal symmetry is not present.
The ferromagnetic (FM) metals that we study here fall into this category and the resulting  critical transition temperature should be seen as an upper bound.   

We present in Table~\ref{tab-lambda} the results for the FM and non-magnetic (NM) cases for both iron and nickel.
\begin{table}[b]
    \centering
    \begin{tabular}{c c c c c c c}
    \hline\hline
         Compound & Magnetism & $\lambda$  & $\lambda^\uparrow$ & $\lambda^\downarrow$ & $\omega_\mathrm{log} [\rm K]$ & $T_{\rm c}^{\rm AD} [\rm K]$ \\ 
         \hline
        Fe (BCC) & FM & 0.2961 & 0.2180 & 0.0781 & 311 & 0.02 \\
        Ni (FCC) & FM & 0.2767 & 0.0040 & 0.2727 & 291 & 0.01\\
        Ni (FCC) & NM &  0.2606  & 0.1303 & 0.1303 & 259 & 0.00 \\
        Fe (BCC) & NM & 1.9124 & 0.9562 & 0.9562 & 28 & 3.64\\
 \hline\hline
    \end{tabular}
    \caption{Total and spin-resolved electron phonon coupling strength $\lambda$ for ferromagnetic (FM) and non magnetic (NM) Fe and Ni.
    The characteristic frequency $\omega_\mathrm{log}$ and Allen-Dynes critical temperature $T_\mathrm{c}^\mathrm{AD}$ are obtained with $\mu^* = 0.13$.
    }
    \label{tab-lambda}
\end{table}
As discussed in the manuscript, the electron-phonon coupling constant $\lambda$ presents a pronounced difference between the two spin channels in Fe and Ni when considering the FM phase. 
If we compare the NM phase with the FM phase, we find that for the case of Ni FCC the value of $\lambda$ is not much affected by magnetic order and only the relative importance of each spin channel is changed. 
This suggests a predominant role of Fermi surface nesting and the number of available states close to the Fermi level. 
On the contrary, for the case of Fe, the values are different with a large $\lambda \approx 1.9$ in the NM case. 
This is due to the low frequency phonons obtained in the NM case, further highlighting the dangers of using phonon dispersions with unstable modes. 
In what concerns the characteristic frequency of the system $\omega_\mathrm{log}$, the values are rather similar between the 4 cases presented in Table~\ref{tab-lambda} with $\omega_\mathrm{log} \approx 300-250\mathrm{K}$ with the exception of NM Fe BCC which is an order of magnitude smaller due to the mentioned near zero frequency modes.

At this point, one can take this characteristic frequency in the Allen-Dynes formula and the compute $\lambda$ to estimate the $T_c$ of an hypothetical $s$-wave superconductor. 
In doing so, we need to provide a reasonable value for the semi-empirical Coulomb potential $\mu^*$. 
Typical values range between 0.1 and 0.2 in the literature~\cite{Kawamura2020,Lilia2022,Bercx2025} and we choose $\mu^* = 0.13$ in Table~\ref{tab-lambda} for illustrative purposes, finding vanishingly small values for all cases, except NM Fe which provides a $T_\mathrm{c}$ of nearly $4\mathrm{K}$ due to the artificially large $\lambda$. 
For the other three cases, it is interesting to find that a phonon mediated mechanism seems rather unfavorable. 
These results are interesting since neither Ni nor Fe have a superconducting transition under ambient conditions and only Fe may show a superconducting transition at very low temperature $T_c \leq 2$~K under high pressure~\cite{Shimizu2001,Jaccard2002}. 
The most common argument to explain this situation is that the long range magnetic order prevents it. 
However, the results given in Table~\ref{tab-lambda} may suggest that the absence of superconductivity on these compounds might be more intrinsic rather than being suppressed by the internal magnetic field.

\section{Momentum convergence of the resistivity}
\begin{figure}[h]
    \centering
    \includegraphics[width=0.99\linewidth]{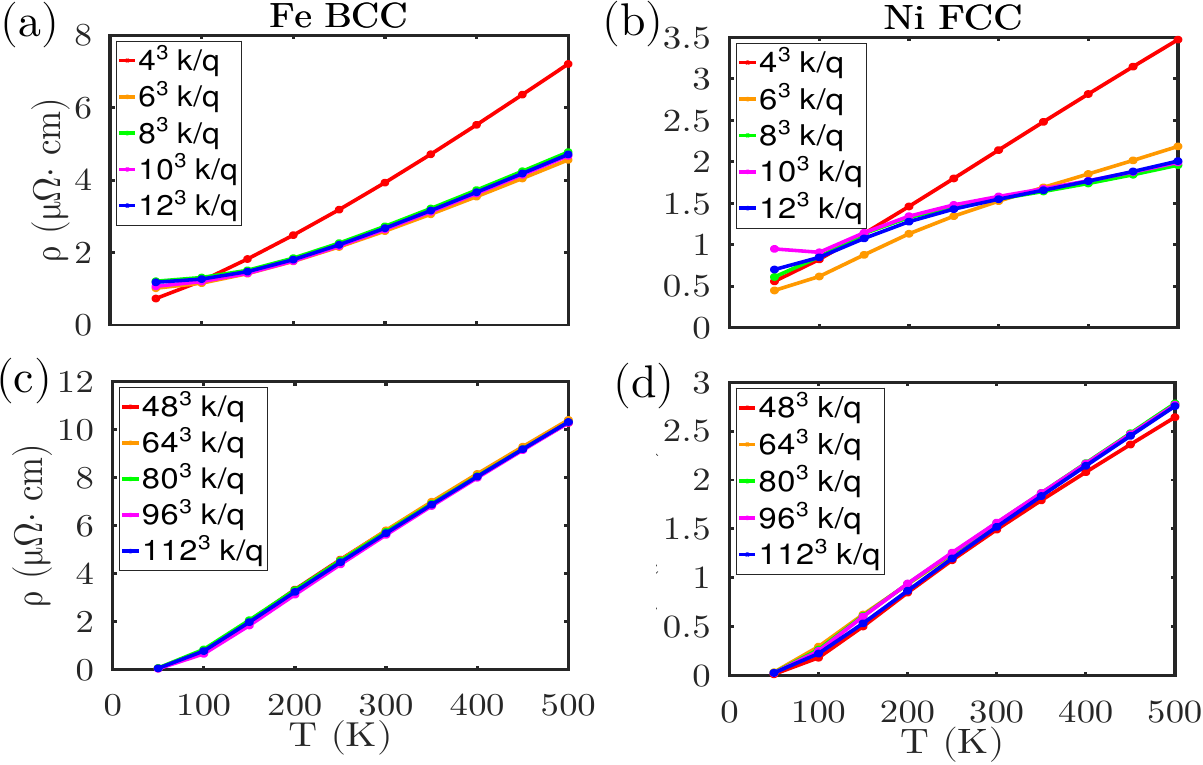}
    \caption{\label{coarse-fine-grid}
    Resistivity using LDA with different $\mathbf{k/q}$ coarse and fine grids for ferromagnetic Fe (a, c), and for ferromagnetic Ni (b, d).} 
\end{figure}

We first converge the coarse $\mathbf{k/q}$ grids using a fixed fine interpolated $30\times30\times30$ grid and show the results in Fig.~\ref{coarse-fine-grid}(a) and (b) for Fe and Ni, respectively.
We find that a $8\times8\times8$ coarse $\mathbf{k/q}$ grid converges below $5\%$ and offers good trade-off between computational cost and precision for both materials. 
With these fixed coarse grids, we repeat the process but with the fine mesh and report the results in Figs.~\ref{coarse-fine-grid}(c) and (d) for Fe and Ni, respectively. 
A convergence below 5\% at 50 K and below 1\% above 150 K is obtained with a fine mesh of $112\times112\times112$ $\mathbf{k/q}$ mesh for both compounds.
We note that $64\times64\times64$ grids are already accurate and provide a maximum deviation of  0.2~$\mu\Omega\cdot$cm.
When not explicitly reported, all resistivity results are obtained with a $8\times8\times8$ coarse $\mathbf{k/q}$ grid  and a fine $112\times112\times112$ $\mathbf{k/q}$ grid.  

\section{Electron band structure with PBE}
\begin{figure}[h]
    \centering \includegraphics[width=0.9\linewidth]{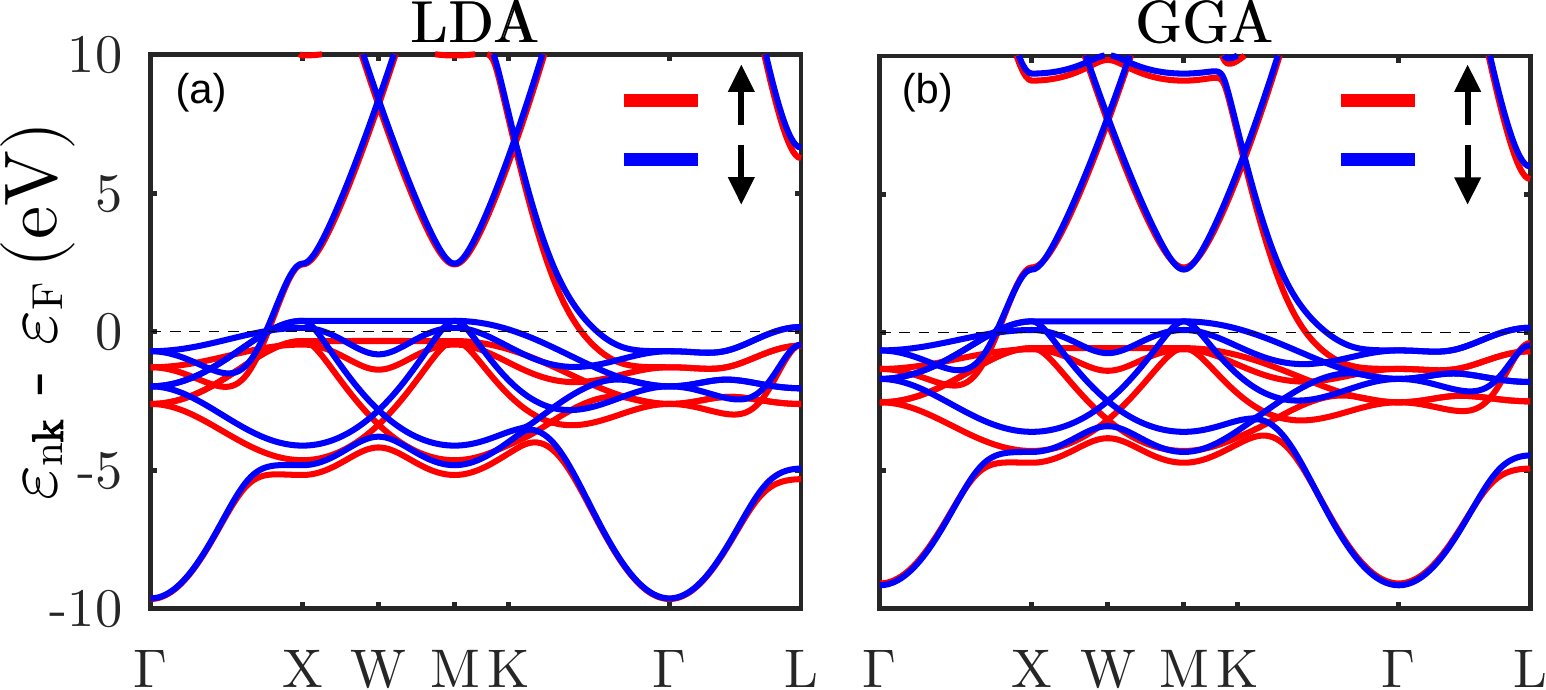}
    \includegraphics[width=0.9\linewidth]{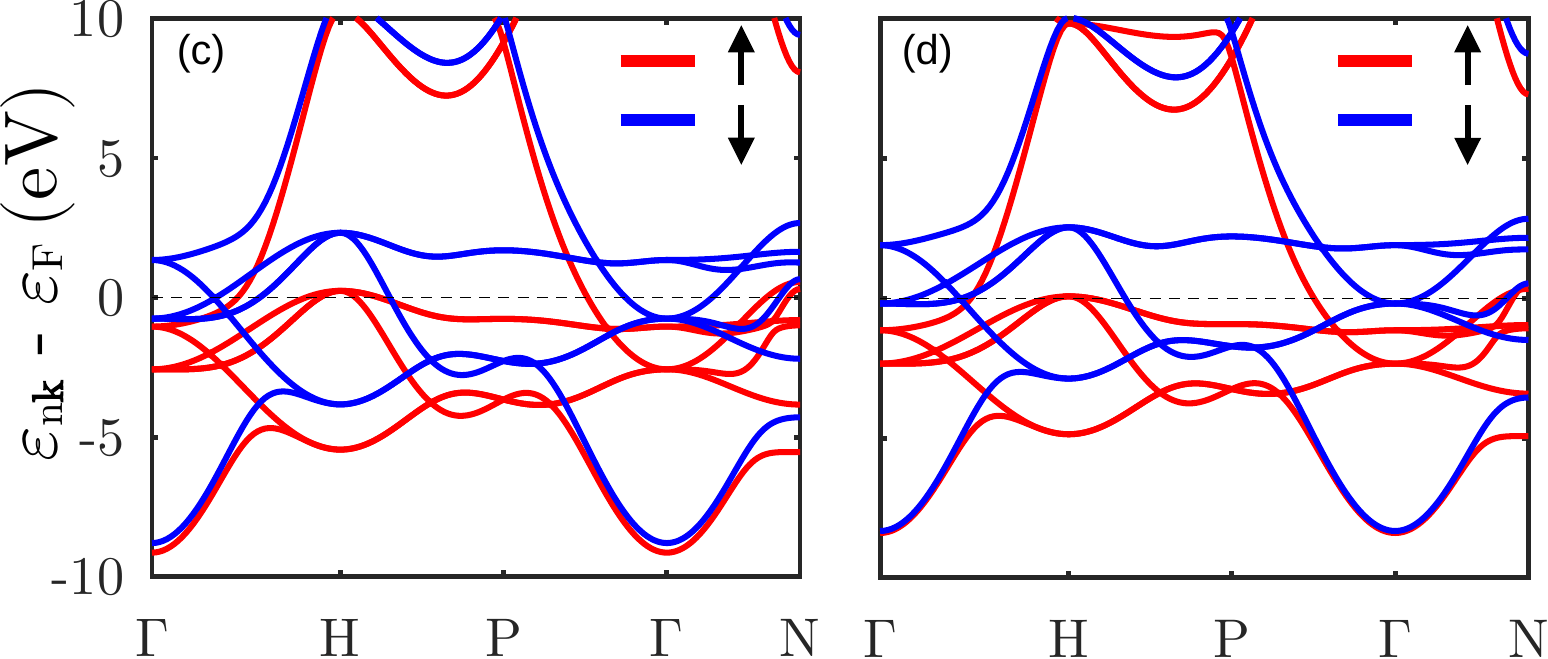}
    \caption{\label{Bands}
    Electronic band structure using \textsc{Quantum Espresso} for ferromagnetic Ni (a) and (b) with LDA and PBE functionals respectively, and for ferromagnetic Fe (c) and (d) with LDA and PBE functionals, respectively. 
    Red solid lines correspond to spin up or spin majority channel. 
    Blue solid lines correspond to spin down or spin minority channel.}
\end{figure}

We compare the electronic band structure of Ni and Fe for LDA and PBE functionals in Fig.~\ref{Bands}.
We find that the exchange splitting between the two channels is enhanced when using PBE instead of LDA. The results shown in Figure~\ref{Bands} for both compounds show a change of the band velocity in both cases when using PBE with respect to LDA. 
\section{Ferromagnetic $\mathrm{Co}$ HCP}

\begin{figure}[h]
    \centering
    \includegraphics[width=0.99\linewidth]{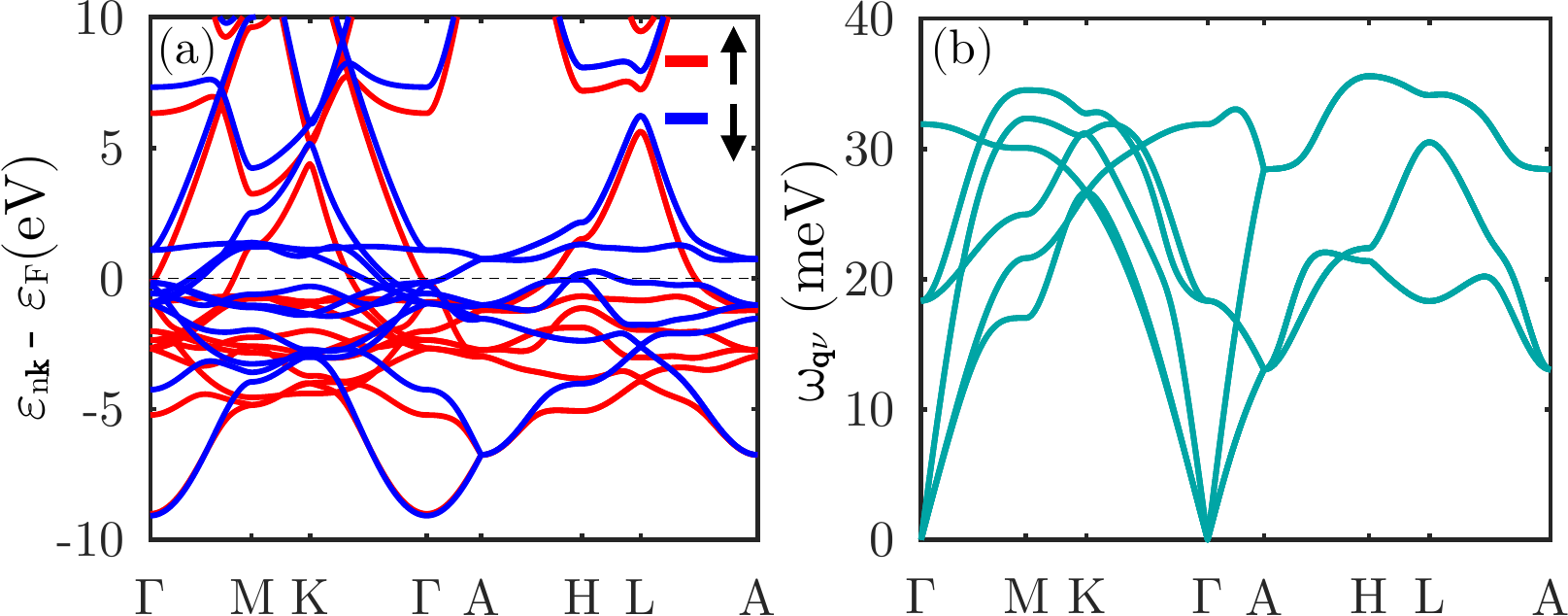}
    \caption{Dispersion relation for (a) electrons and (b) phonons in ferromagnetic Co hcp.}
    \label{Co-bands}
\end{figure}

Indicatively, we also compute the electronic and phonon band structure of ferromagnetic Co hcp with the PBE functional showing the results in Fig.~\ref{Co-bands}. 
For the electronic bandstructure, a $24\times24\times12$ $\mathbf{k}$ grid is used, and for the phonons a $8\times 8\times 4$ $\mathbf{q}$ grid is used.

\begin{figure}[h]
    \centering
    \includegraphics[width=0.75\linewidth]{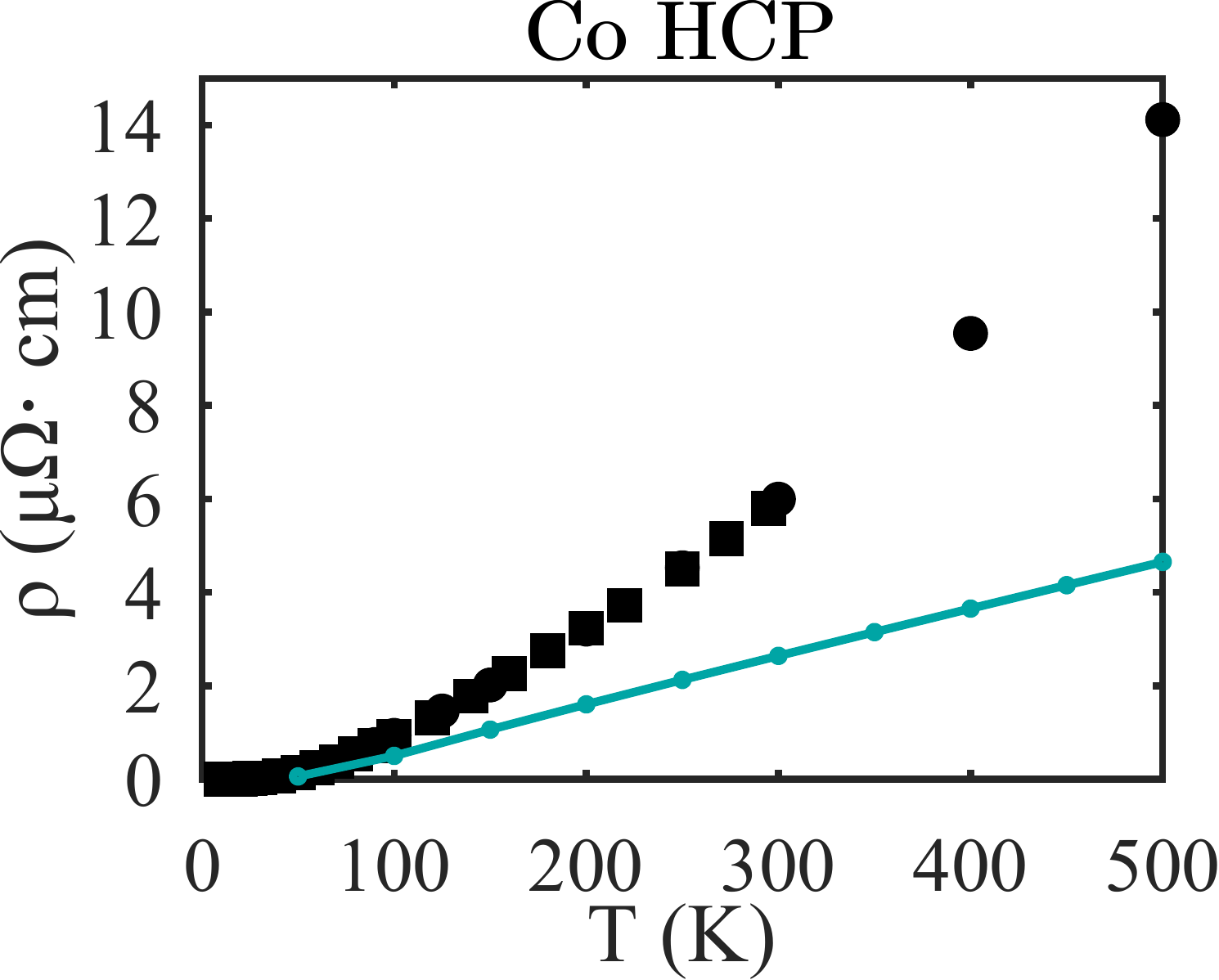}
    \caption{\label{Co-res}
    Electrical resistivity of ferromagnetic Co.
    Solid lines are the calculated resistivity using PBE.
    Black symbols represent the experimental resistivity from Refs.~\cite{1959,Laubitz1976}.}
\end{figure}

Furthermore, we compute the electrical resistivity with an $8\times8\times4$ coarse and a $96\times96\times48$ fine $\mathbf{k/q}$ grids and compare it with the experiments in Fig.~\ref{Co-res}.
We find that up to 300~K, the electron-phonon interactions can account for up to 50\% of the experimental resistivity placing Co in between the behavior of Fe and Ni discussed in the manuscript.

\section{Non-magnetic solution for $\rm Fe$ and $\rm Ni$}

\begin{figure}[b]
    \centering
    \includegraphics[width=0.99\linewidth]{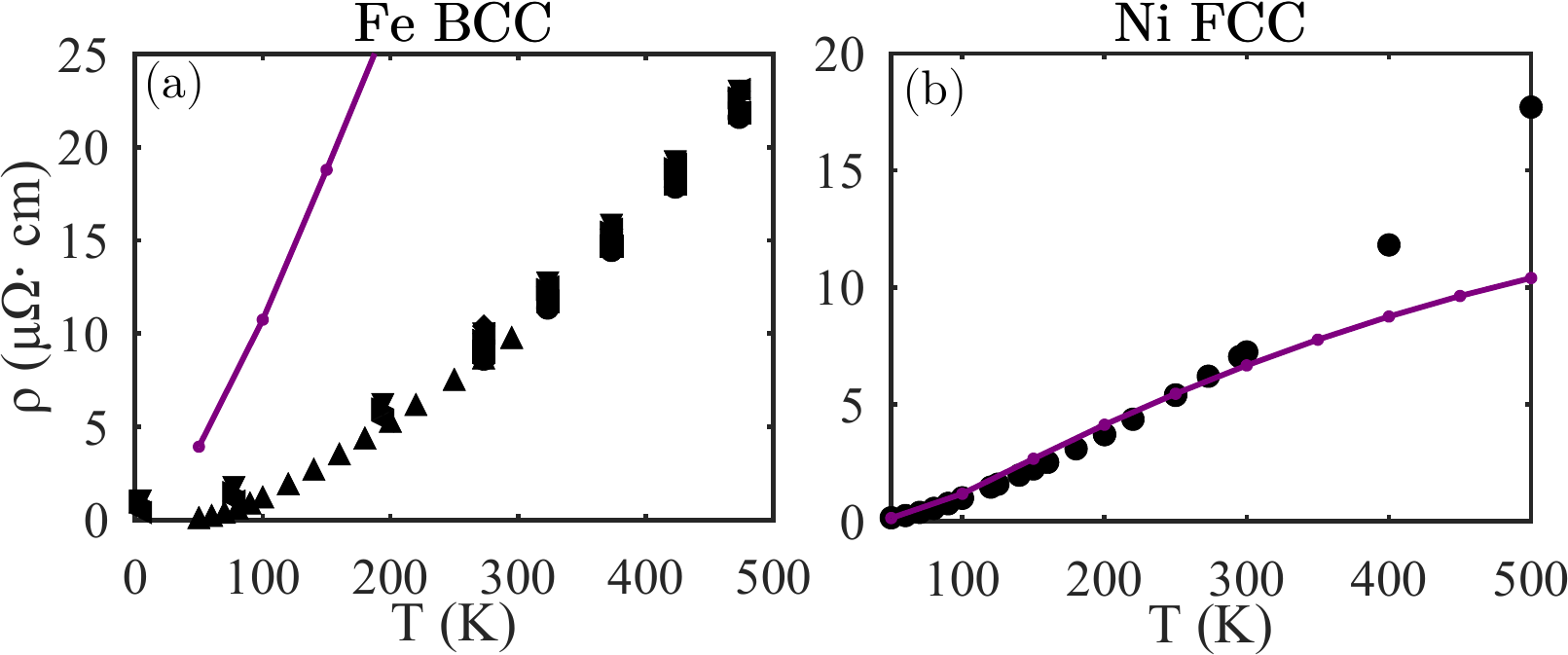}
    \caption{Electrical resistivity of non-magnetic Fe (a), and  Ni (b).
    Solid lines are the calculated resistivity using LDA Black symbols represent the experimental resistivity from Refs.~\cite{1959,Laubitz1976,Fulkerson1966}.}
    \label{Res-lda-NM}
\end{figure}

Finally, we also show in Fig.~\ref{Res-lda-NM} the electrical resistivity for non-magnetic Fe and Ni using LDA and compared with experimental results.
When the phonons are soft (negative), we neglect their contribution to the resistivity. 
In the case of Fe in Fig.~\ref{Res-lda-NM}(a), we can clearly see a generalized overestimation of the resistivity similar to the case with a PBE  functional. 
This behavior clearly shows the need to include magnetism in Fe. 
In the case of Ni in Fig.~\ref{Res-lda-NM}(b), we obtain a resistivity curve that below 300~K slightly overestimates the resistivity and then above 300~K it starts to underestimate it. 
Although one may be tempted to think that this is something satisfactory, since below 300~K the resistivity curve is quite close to the experimental values, one should note that both the electronic dispersion and the phonon dispersion change significantly between LDA and PBE.
In particular, the phonon dispersion of LDA clearly overestimates the phonon frequencies in contrast with the PBE calculation which yields a rather good match with the experiments. 
Thus, this better agreement might be due to error cancelations.

\bibliography{bibliography-SI}